\documentclass[aps,prd,preprintnumbers,superscriptaddress,nofootinbib,onecolumn,notitlepage]{revtex4-1}
\usepackage{ulem}
\usepackage[dvipdfmx]{graphicx}
\usepackage{bm,latexsym,amsmath,amssymb,amsfonts}
\usepackage[colorlinks=true,linkcolor=magenta,citecolor=magenta]{hyperref}
\usepackage{color}
\usepackage{framed}

\newcommand*{\ba}{\begin{eqnarray}}
\newcommand*{\ea}{\end{eqnarray}}

\newcommand{\calC}{{\cal C}}

\newcommand{\calE}{{\cal E}}
\newcommand{\calH}{{\cal H}}
\newcommand{\calK}{{\cal K}}
\newcommand{\calL}{{\cal L}}
\newcommand{\calM}{{\cal M}}
\newcommand{\calN}{{\cal N}}
\newcommand{\calR}{{\cal R}}

\newcommand{\dd}{{\rm d}}

\newcommand*{\Stuckelberg}{St\"uckelberg~}

\newcommand{\calCt}{\widetilde{\cal C}}
\newcommand{\lambdat}{\widetilde{\lambda}}

\newcommand{\Bhi}{B^h_i}
\newcommand{\Bfi}{B^f_i}
\newcommand{\Fhi}{F^h_i}
\newcommand{\Ffi}{F^f_i}

\newcommand{\piBhi}{\pi_{B^h_i}}
\newcommand{\piBfi}{\pi_{B^f_i}}
\newcommand{\piFhi}{\pi_{F^h_i}}
\newcommand{\piFfi}{\pi_{F^f_i}}

\begin{document}

\title{
On Lorentz-invariant bi-spin-2 theories
}

\author{Rampei~Kimura}
\affiliation{Waseda Institute for Advanced Study, Waseda University, 1-6-1 Nishi-Waseda, Shinjuku, Tokyo 169-8050, Japan}

\author{Atsushi~Naruko}
\affiliation{Center for Gravitational Physics, Yukawa Institute for Theoretical Physics, Kyoto University, Kyoto 606-8502, Japan}

\author{Daisuke~Yamauchi}
\affiliation{Faculty of Engineering, Kanagawa University, Kanagawa-ku,
	Yokohama-shi, Kanagawa, 221-8686, Japan}

\preprint{YITP-20-144}

\begin{abstract}
We investigate a Lorentz invariant action which is quadratic in two rank-2 symmetric tensor fields in Minkowski spacetime. We apply a scalar-vector-tensor decomposition to two tensor fields by virtue of 3-dimensional rotation-invariance of Minkowski spacetime and classify theories with seven degrees of freedom based on the Hamiltonian analysis. We find two new theories, which cannot be mapped from the linearized Hassan-Rosen bigravity. In these theories, the new mass interactions can be allowed thanks to the transverse diffeomorphism invariance of action.
\end{abstract}

\maketitle

\section{Introduction}

The attempt to seek ghost-free massive gravity theories has  again attracted considerable attention by the discovery of de Rham-Gabadadze-Tolley (dRGT) massive gravity \cite{deRham:2010ik}. The first attempt of constructing massive spin-2 theory has been carried out by Fierz and Pauli, and it is the quadratic action for a massive spin-2 particle in a flat spacetime \cite{Fierz:1939ix}. Once we embed this into a curved spacetime, the behavior of the massive spin-2 field does not smoothly connect to the well-known massless one, i.e., the linearized general relativity \cite{Zakharov:1970cc,vanDam:1970vg}. The discontinuity found by van Dam, Veltman, and Zhakalov turned out to be an artifact of the truncation at linear order, and the massive spin-2 theory in fact has the continuous massless limit when taking into account nonlinearities as pointed out by Vainshtein \cite{Vainshtein:1972sx}. Nonetheless, an unwanted degree of freedom (DOF),  Boulware-Deser ghost \cite{Boulware:1972aa}, which is absent at linear order, reappears at nonlinear level, and it unfortunately behaves as Ostrogradsky's ghost \cite{Ostrogradsky1850}. In dRGT massive gravity, such an unwanted degree of freedom is successfully eliminated by the careful choice of nonlinear potential terms \cite{deRham:2010ik,deRham:2010kj}. Although the dRGT massive gravity possess the cosmological constant solution in a cosmological 
 background \cite{Gumrukcuoglu:2011ew}, it is perturbatively unstable \cite{DeFelice:2012mx,Gumrukcuoglu:2011zh}.
For this reason, one needs to seek a ghost-free extension of massive gravity which 
should be at least cosmologically viable and stable. 
Such an attempt without introducing an extra DOF has been investigated, taking into account derivative interactions \cite{Hinterbichler:2013eza,Kimura:2013ika,deRham:2013tfa} and metric transformation \cite{Gumrukcuoglu:2019rsw}, but most of them are not successful unfortunately. 
Recently, by breaking the translation invariance of the \Stuckelberg field, new extended theories of massive gravity have found, and their cosmological perturbations are stable around cosmological backgrounds  \cite{deRham:2014lqa,Kenna-Allison:2019tbu,Gumrukcuoglu:2020utx}. 

Another way to extend massive gravity is to introduce the second
dynamical symmetric tensor field. In massive gravity theories,
to give mass to graviton, in addition to the metric $g_{\mu\nu}$, one needs to introduce the so-called reference metric $f_{\mu\nu}$, which is usually taken to be a Minkowski metric. In massive bigravity theories, the reference metric can be promoted to be a dynamical tensor field by introducing its kinetic terms. The simplest extension of dRGT massive gravity is proposed by Hassan and Rosen by adding the Einstein-Hilbert kinetic terms even for the second metric \cite{Hassan:2011zd}. In Hassan-Rosen bigravity, the total number of physical DOFs is seven, which consists of two from a massless graviton and five from a massive graviton. This fact can be easily seen by expanding both metric around Minkowski spacetime, that is, $g_{\mu\nu} \to \eta_{\mu\nu} + h_{\mu\nu}/M_g$ and $f_{\mu\nu} \to \eta_{\mu\nu} + f_{\mu\nu}/M_f$, where 
$M_g$ and $M_f$ are respectively the Planck mass for the metric $g_{\mu\nu}$ and $f_{\mu\nu}$. 
Then the quadratic Lagrangian is given by \cite{Hassan:2011zd}
\ba
{\cal L}^{\rm (2)}_{\rm HR}&=&-\left( h_{\mu\nu}\widehat{\cal E}^{\mu\nu\alpha\beta}h_{\alpha\beta}+f_{\mu\nu}\widehat{\cal E}^{\mu\nu\alpha\beta}f_{\alpha\beta}\right)
-\frac{m^2M_{\rm eff}^2}{4}\Biggl[\left(\frac{h^\mu{}_\nu}{M_g}-\frac{f^\mu{}_\nu}{M_f}\right)^2-\left(\frac{h}{M_g}-\frac{f}{M_f}\right)^2\Biggr]
\,.
\ea
Here $\widehat{\cal E}^{\mu\nu\alpha\beta}$ is the linearized Einstein-Hilbert kinetic operator defined as 
\ba
\widehat{\cal E}^{\mu\nu}_{~~\alpha\beta}
= 
\left[ \eta^{(\mu}_{~\alpha}\eta^{\nu)}_{~\beta}-\eta^{\mu\nu}\eta_{\alpha\beta}\right] \square
-2 \partial^{(\mu}\partial_{(\alpha}\eta^{\nu)}_{~\beta)}
+\partial^\mu \partial^\nu \eta_{\alpha\beta} + \partial_\alpha \partial_\beta \eta^{\mu\nu} \,,
\label{action:HR}
\ea	
 where the round brackets denote the symmetrization of indices, $m$ is the mass of graviton, and the effective Planck mass is given by $M_{\rm eff}^2 = (1/M_g^2 + 1/M_f^2)^{-1}$.
The mixing terms between $h$ and $f$ in the mass terms can be removed by introducing the linear combination of two metric,  
\ba
\frac{1}{M_{\rm eff}}u_{\mu\nu}\equiv\frac{1}{M_f}h_{\mu\nu}+\frac{1}{M_g}f_{\mu\nu}
\,, \qquad\frac{1}{M_{\rm eff}}v_{\mu\nu}\equiv\frac{1}{M_f}h_{\mu\nu}-\frac{1}{M_g}f_{\mu\nu}
\,.
\ea
Then the Lagrangian becomes
\ba
{\cal L}&=&-\left( u_{\mu\nu}\widehat{\cal E}^{\mu\nu\alpha\beta}u_{\alpha\beta}+v_{\mu\nu}\widehat{\cal E}^{\mu\nu\alpha\beta}v_{\alpha\beta}\right)
-\frac{m^2}{4}\left( v^{\mu\nu}v_{\mu\nu}-v^2\right)
\,.
\label{action:HR-diagonal}
\ea
This clearly shows that Hassan-Rosen bigravity at linear order consists of the lienarized general relativity for $u_{\mu\nu}$ and the Fierz-Pauli theory for $v_{\mu\nu}$. The absence of the Boulware-Deser 
ghost has been proved in \cite{Hassan:2011ea}.

Since the construction of bigravity theory is inspired by the dRGT theory, it is not trivial whether the mass interaction of Hassan-Rosen bigravity is unique or not. As for the uniqueness of dRGT mass term in linear massive gravity theories, see also 
\cite{Naruko:2018akp}. For this reason, one might be able to find a new type of mass interactions in bimetric gravity theories. However, such a construction would be extremely difficult to start with a curved spacetime. To this end, in this paper, we investigate 
a theory with bi-spin-2 particle in a flat spacetime, which could represent the linear expansion of a certain nonlinear massive bigravity. 

This paper is organized as follows. In Sec.~II, we give an action for two rank-$2$ tensor fields in our setup and decompose them into scalar, vector, and tensor sectors based on transformation properties of tensors with respect to a $3$-dimensional spatial rotation. In
 Sec.~III, we give ghost-free conditions for the tensor mode. In Sec.~IV, we perform Hamiltonian analysis and derive the conditions to have $2$ physical DOFs in the vector sector. In Sec.~V, we investigate the scalar sector and classify theories with $1$ scalar DOFs. Sec. VI is devoted to the summary. In Appendix A, we introduce a linear field redefinition and investigate the reduction of the model parameter space. In Appendix B, we provide an explicit expression of the Lagrangian in the scalar sector. In Appendix C, we perform the Hamiltonian analysis of the vector sector with $2$ primary constraints. In Appendix D, we investigate the scalar sector with $2$ primary constraints.

\section{Setup}
In the present paper, we consider a Lorentz invariant action for two rank-2 symmetric tensor fields,  $h_{\mu\nu}$ and $f_{\mu\nu}$, and consider the most general quadratic action  which contains up to two derivatives with respect to spacetime for each term\footnote{Strictly speaking, one can also include a Lorentz-invariant scalar $ {\eta_{\mu\nu}x^\mu x^\nu= -t^2 + {\bf x}^2}$ for theories invariant under a  global Lorentz transformation. Once introducing this scalar quantity, the analysis would be more complicated. For simplicity, we here do not consider such a possibility.}. In general, these symmetric tensor fields possess 20 DOFs in total, and therefore we should impose some conditions to eliminate unwanted DOFs, which could behave as ghost. Due to the complexity of the analysis, we only focus on theories with 7 physical DOFs,
namely $2 \times 2$ (tensor) + $2$ (vector) + $1$ (scalar) DOFs, as in the Hassan-Rosen bigravity \cite{Hassan:2011zd} that consists of massless and massive spin-2 fields in the linearized limit. 
As preparation for later analysis, in this section, we introduce the generic action for bi-spin-$2$ tensor field and scalar-vector-tensor decomposition of it. 
For the Hamiltonian analysis in Fourier space, we follow the procedure developed in \cite{Sugano1982,Sugano1986,Sugano:1989rq} and adopt the notation in \cite{Naruko:2018akp}.

\subsection{Double spin-2 theory}
Let us consider a generic action for two rank-2 symmetric tensor fields up to the quadratic order in fields around Minkowski spacetime,
\begin{align}
 S &= \int \dd^4 x \Bigl(
 - \calK_h^{\alpha \beta | \mu \nu \rho \sigma} h_{\mu \nu, \alpha} h_{\rho \sigma, \beta}
  - \calK_f^{\alpha \beta | \mu \nu \rho \sigma} f_{\mu \nu, \alpha} f_{\rho \sigma, \beta}
 -{{\cal G}^{\alpha \beta | \mu \nu \rho \sigma}} h_{\mu \nu, \alpha} f_{\rho \sigma, \beta}\notag\\
 & \qquad \qquad \qquad
 - \calM_h^{\mu \nu \rho \sigma} h_{\mu \nu} h_{\rho \sigma} 
- \calM_f^{\mu \nu \rho \sigma} f_{\mu \nu} f_{\rho \sigma} 
 - \calN^{\mu \nu \rho \sigma} h_{\mu \nu} f_{\rho \sigma} \Bigr) \,,
 \label{action}
\end{align}
where the coefficients $\calK_\Phi$, ${\cal G}$, $\calN$, and $\calM_\Phi$ consist of 
all the possible combinations with the Minkowski metric $\eta_{\mu\nu}$,  
\begin{align}
 \calK_\Phi^{\alpha \beta | \mu \nu \rho \sigma}
 &= \kappa_{\Phi1} \eta^{\alpha \beta} \eta^{\mu \rho} \eta^{\nu \sigma}
 + \kappa_{\Phi2} \eta^{\mu \alpha} \eta^{\rho \beta} \eta^{\nu \sigma}
 + \kappa_{\Phi3} \eta^{\alpha \mu} \eta^{\nu \beta} \eta^{\rho \sigma}
 + \kappa_{\Phi4} \eta^{\alpha \beta} \eta^{\mu \nu} \eta^{\rho \sigma} \,,\\ 
 {\cal G}^{\alpha \beta | \mu \nu \rho \sigma}
&= 
(l_1 \eta^{\alpha \beta} \eta^{\mu \rho}
+ l_2 \eta^{\mu \alpha} \eta^{\rho \beta}) \eta^{\nu \sigma}
+ (l_3 \eta^{\alpha \mu} \eta^{\nu \beta}
+ l_4 \eta^{\alpha \beta} \eta^{\mu \nu}) \eta^{\rho \sigma} 
+ l_5 \eta^{\mu \nu} \eta^{\beta \sigma}
\eta^{\alpha \rho}  \,,
\\ 
 \calM_\Phi^{\mu \nu \rho \sigma}
  &= \mu_{\Phi1} \eta^{\mu \rho} \eta^{\nu \sigma}
 + \mu_{\Phi2} \eta^{\mu \nu} \eta^{\rho \sigma} \,, \\
\calN^{\mu \nu \rho \sigma}
&= n_1 \eta^{\mu \rho} \eta^{\nu \sigma}
+ n_2 \eta^{\mu \nu} \eta^{\rho \sigma} \,,
\end{align}
and we defined the label $\Phi = (h, f)$. 
A comma denotes a partial derivative with respect to coordinates.
Here, $\kappa_{\Phi1,\Phi2,\Phi3,\Phi4}$, $l_{1,2,3,4,5}$, $\mu_{\Phi1,\Phi2}$, and $n_{1,2}$ are constant parameters. 
The linearized Hassan-Rosen bigravity corresponds to 
\ba
&&\kappa_{h2}=-\kappa_{h3}=2\kappa_{h4}=-2\kappa_{h1} \,, \qquad 
\kappa_{f2}=-\kappa_{f3}=2\kappa_{f4}=-2\kappa_{f1}\,, \qquad l_1=l_2=l_3=l_4=l_5=0\,, \notag\\
&&\mu_{h2}=-\mu_{h1} \,, \qquad
\mu_{f2}=-\mu_{f1} \,, \qquad
n_2=-n_1=2\sqrt{\mu_{h1}\mu_{f1}} \,,
\label{HR1}
\ea
 as shown in \eqref{action:HR} and this theory is invariant under the gauge transformation,
\ba
h_{\mu\nu} \to 
h_{\mu\nu} + {1 \over 2\sqrt{\mu_{h1}}}(\partial_\mu \xi_\nu+ \partial_\nu \xi_\mu)\,, \qquad
f_{\mu\nu} \to 
f_{\mu\nu} + {1 \over 2\sqrt{\mu_{f1}}}(\partial_\mu \xi_\nu+ \partial_\nu \xi_\mu)\,.
\ea
Alternatively, one can diagonalize the mass terms to remove $n_1$ and $n_2$ without changing the kinetic terms by taking linear combinations of $h_{\mu\nu}$ and $f_{\mu\nu}$, and then the resultant theory satisfies
\ba
&&\kappa_{h2}=-\kappa_{h3}=2\kappa_{h4}=-2\kappa_{h1} \,, \qquad 
\kappa_{f2}=-\kappa_{f3}=2\kappa_{f4}=-2\kappa_{f1}\,,\ \notag\\
&&\mu_{h2}=-\mu_{h1} \,, \qquad 
\mu_{f1}=\mu_{f2}=0 \,, \qquad
l_1=l_2=l_3=l_4=l_5=n_1=n_2=0 \,,
\label{HR2}
\ea
 as found in \eqref{action:HR-diagonal}.
Then this theory with \eqref{HR2} is invariant under the gauge transformation
\ba
h_{\mu\nu} \to  
h_{\mu\nu} \,, \qquad
f_{\mu\nu} \to 
f_{\mu\nu} +\partial_\mu \xi_\nu+ \partial_\nu \xi_\mu\,.
\ea
Thus, this is nothing but the Fierz-Pauli massive spin-2 field for $h_{\mu\nu}$ and the linearized general relativity for $f_{\mu\nu}$.

\subsection{Scalar-vector-tensor decomposition}
Following \cite{Naruko:2018akp}, we decompose the rank-2 symmetric tensor fields $h_{\mu\nu}$ and $f_{\mu\nu}$ into 
transverse-traceless tensors, transverse vectors, and scalars as 
\begin{align}
	\Phi_{0 0} &= \Phi^{0 0} = - 2 \alpha_\Phi \,, \qquad
	\Phi_{0 i} = - \Phi^{0 i} ={\beta^\Phi_{, i}}  
		+ B^\Phi_{ i} \quad (B^i{}_{\Phi, i} = 0)\,, \\
	\Phi_{i j} &= \Phi^{i j}
	= 2 {\cal R}_\Phi \delta_{i j} + 2\, {{{\cal E}}^\Phi_{, i j}} 
		+ F^\Phi_{ i \,, j} + F^\Phi_{ j \,, i}
	+ 2 H^\Phi_{ i j} \quad ( F^i{}_{\Phi, i} = 0 \,, \quad H^i{}_{\Phi i} = H^{i j}{}_{\Phi, j} = 0) \,.
\end{align}
Here, scalar, vector and tensors are defined based on transformation properties with respect to a $3$-dimensional rotation in Minkowski spacetime, and the transverse-traceless tensors ${H^\Phi_{ij}}$, two transverse vectors $B^\Phi_{ i}$ and ${F^\Phi_{i}}$, and four scalars $\alpha_\Phi, {\beta_\Phi}, {\cal R}_\Phi$, and ${{\cal E}_\Phi}$ respectively have two, four, and four components in each $\Phi$. 
Since we focus on theories with 7 DOFs, to be more precise $2 \times 2$ (tensor) + $2$ (vector) + $1$ (scalar) DOFs, we need to eliminate 6 components of the transverse-vectors and 7 components of the scalars, and then the final DOFs becomes $20-6-7 = 7$ DOFs. 
Under this decomposition, the quadratic action can be always separated into three parts 
which solely consists of scalar, vector, and tensor perturbations respectively 
:
\begin{align}
	S [h_{\mu \nu}, f_{\mu\nu}]
	&= S^S [\alpha_\Phi \,, {\beta_\Phi} \,, {\cal R}_\Phi \,, {{\cal E}_\Phi}\,] + S^V[{B^\Phi_{i}} \,, {F^\Phi_{i}}] + S^T [{H^\Phi_{i j}}] \,.
\end{align}
In the following section, we will examine each sector and derive conditions to have theories with $7$ DOFs by the Hamiltonian analysis. Hereafter, we replace all the spatial derivatives as $\partial^2 \to -k^2$ after integrating by parts, where $k$ is the wavenumber in the Fourier space.

\section{tensor sector}
The action in the tensor sector is found to be
\ba
S^T[H_{ij}^h, H_{ij}^f] &=&  4\int dt d^3k   \Big[
\kappa_{h1} ({\dot H}^h_{ij} )^2 -(\kappa_{h1} k^2 + \mu_{h1})(H^h_{ij} )^2 
+\kappa_{f1} ({\dot H}^f_{ij} )^2 -(\kappa_{f1} k^2 + \mu_{f1})(H^f_{ij} )^2 \nonumber\\
&&~~~~~~~~~~~~~~~
{-  l_1 {\dot H}_{ij}^h {\dot H}_f^{ij}
-  (k^2 l_1 + n_1) H_{ij}^h H_f^{ij}}
\Big] \,,~~
\ea
where a dot denotes the time derivative. It is manifest that the action is symmetric under the replacement $h$ and $f$, and hence the result will be applied to both modes in parallel.
Throughout this paper, assuming $\kappa_{h1} \neq 0$, we set  
\ba
l_1 =0\,,
\ea 
which can be achieved by a field redefinition of $f$ without loss of generality
(See Appendix \ref{sec:field redef}). 
Thanks to $l_1 =0$ by the field redefinition, the kinetic matrix composed of $h$ and $f$ is diagonal, and the existence and ghost-free conditions of both the tensor modes requires 
 \ba
 \kappa_{h1} > 0\,, \qquad \kappa_{f1} > 0 \,.
 \label{ConditionT}
 \ea
Hereafter, we impose the condition \eqref{ConditionT}, and it is manifest that the physical degrees of freedom in the tensor sector is two for each field.

\section{Vector sector}
In this section, we perform the Hamiltonian analysis for the vector variables. In order to have a theory with $7$ DOFs in total, the vector sector should have $2$ physical DOFs, which means the reduction of the phase space is necessary in the view point of the Hamiltonian analysis. 
We first rescale {$F^\Phi_i$} as {$F^\Phi_i \to F^\Phi_i/k$} for convenience.  Then, the action in the vector sector is given by
\begin{align}
S^V [{B^\Phi_{i}}, {F^\Phi_{i}}] = \int dt d^3k \, \Big( \calL^{V}_{\rm kin} + \calL^{V}_{\rm cross} + \calL^{V}_{\rm mass} \Big) \,,
\end{align}
 where each Lagrangian is given by
\begin{align}
	\calL^{V}_{\rm kin}&=
-(2\kappa_{h1}+\kappa_{h2}) ({\dot B}^h_i)^2 
+ 2\kappa_{h1} ({\dot F}^h_i)^2
-(2\kappa_{f1}+\kappa_{f2}) ({\dot B}^f_i)^2 + 2\kappa_{f1} ({\dot F}^f_i)^2
-l_2 {\dot B}^h_i {\dot B}^f_i \,,
\label{kinmat-vector}
\\
\calL^{V}_{\rm cross}&=
2k \kappa_{h2}\Bhi {\dot F}^h_i +  2k \kappa_{f2}\Bhi {\dot F}^f_i
+ kl_2(\Bhi {\dot F}^f_i+\Bfi {\dot F}^h_i) \,, \\
 \calL^{V}_{\rm mass} &=
2(k^2 \kappa_{h1}+\mu_{h1}) (\Bhi)^2
-\left(k^2(2 \kappa_{h1} + \kappa_{h2}) +2 \mu_{h1} \right) (\Fhi )^2
+ 2(k^2 \kappa_{f1}+\mu_{f1}) (\Bfi)^2\nonumber\\
&~~~~~~~~~~
-\left(k^2(2 \kappa_{f1} + \kappa_{f2}) +2 \mu_{f1} \right) (\Ffi )^2
+2n_1\Bhi\Bfi -\left(k^2l_2+2n_1\right) \Fhi\Ffi 
\,.
\end{align}
The relation between conjugate momentum $\pi_{\Phi} \equiv \partial \calL / \partial {\dot \Phi}$ and the time derivatives of canonical variables of $B^\Phi_i$ and $F^\Phi_i$ is found to be
\begin{align}
	\begin{pmatrix}
		\piBhi \\ 
		\piBfi \\
		\piFhi \\ 
		\piFfi 
	\end{pmatrix}
	=
	\begin{pmatrix}
		- 2 (2 \kappa_{h 1}+\kappa_{h2})& - l_2 &0&0 \\
		- l_2 & - 2 (2\kappa_{f1} + \kappa_{f2}) & 0 & 0 \\
		0&0& 4\kappa_{h1}&0\\
		0&0&0& 4\kappa_{f1} 
	\end{pmatrix}
	\begin{pmatrix}
		\dot{B}^h_i \\
		\dot{B}^f_i \\
		\dot{F}^h_i\\
		\dot{F}^f_i 
	\end{pmatrix}
+		\begin{pmatrix}
	0& 0 &0&0 \\
	0 & 0& 0 & 0 \\
	2 k \kappa_{h2} & k l_2 & 0&0\\
	k l_2& 2k\kappa_{f2}&0& 0
\end{pmatrix}
\begin{pmatrix}
	\Bhi \\
	\Bfi \\
	\Fhi\\
	\Ffi 
\end{pmatrix} \,.
\end{align}
Then the Hamiltonian is defined by 
\begin{align}
	\calH^V &= {\dot{B}_i^h \pi_{B_i^h}+\dot{B}_i^f \pi_{B_i^f} + \dot{F}_i^h \pi_{F_i^h} + \dot{F}_i^f \pi_{F_i^f}} - \calL^V \,.
\end{align}
As one can see from \eqref{kinmat-vector}, the kinetic parts of $B^\Phi_i$ and $F^\Phi_i$ are completely decoupled, i.e., the kinetic matrix is block diagonalized and the kinetic terms for $F^\Phi_i$ indeed exist
 since we have imposed \eqref{ConditionT}.
Therefore, in order to see the degeneracy of the vector sector, it allows us to consider only the kinetic matrix of $B^\Phi_i$, which is
\begin{align}
	{\cal K}_V
	=
	\begin{pmatrix}
		- 2 (2 \kappa_{h 1}+\kappa_{h2})& - l_2  \\
		- l_2 & - 2 (2\kappa_{f1} + \kappa_{f2}) 
	\end{pmatrix}\,.
\end{align}
The eigenvalue equation ${\cal F}_V(\lambda)$ of the kinetic matrix $ {\cal K}_V$ is found to be
\ba
{{\cal F}_V(\lambda)\equiv\det ({\cal K}_V-\lambda I)
=(4\kappa_{f1} + 2\kappa_{f2} + \lambda)(4\kappa_{h1}+2\kappa_{h2} + \lambda) - l_2^2}
=0\,.
\ea
The determinant of the kinetic matrix is simply given by $\det{\cal K}_V = {\cal F}_V (0)$. 

Now we would like to classify the cases based on the number of primary constraints as follows : 
\ba
&& \textrm{2 primary constraints : }\quad {\cal F}_V(0)=0 \quad \& \quad {\cal F}'_V(0)\neq 0 \quad \longleftrightarrow\quad \kappa_{f2} = -2\kappa_{f1} + \frac{l_2^2}{4 (2\kappa_{h1}+\kappa_{h2})} \,,  \\ 
&& \textrm{4 primary constraints : }\quad {\cal F}_V(0)=0 \quad \& \quad {\cal F}'_V(0)=0 \quad \longleftrightarrow\quad
\kappa_{h2}=-2\kappa_{h1}
\quad \& \quad 
\kappa_{f2}=-2\kappa_{f1}
\quad \& \quad 
l_2=0 \,. \notag\\
\ea
Here, at this point, $h$ and $f$ are symmetric, therefore, the case with $2\kappa_{h1}+\kappa_{h2}=0$ in the case of 2 primary constraints can be obtained by simply replacing $h$ and $f$. 
 When there are only two primary constraints, the Hamiltonian analysis shows that the number of the final physical DOFs can be at least 4, and it is the undesired number. The analysis for two primary case is summarized in the Appendix \ref{app:v2p}. For this reason, hereafter we only consider 4 primary case, where both $B^h_i$ and $B^f_i$ becomes non-dynamical. 

In this case, we have four primary constraints, which are given by 
\ba
\calC^{(1)}_{\Bhi} &\equiv& \piBhi  \approx 0 \,,\\
\calC^{(1)}_{\Bfi} &\equiv& \piBfi  \approx 0 \,, 
\ea
and we define the total Hamiltonian by adding the Lagrange multipliers $ \lambda_{\Bhi} $ and $ \lambda_{\Bfi} $, 
\ba
\calH^V_{T} = \calH^V 
+ \lambda_{\Bhi} \calC^{(1)}_{\Bhi} 
+ \lambda_{\Bfi} \calC^{(1)}_{\Bfi} \,.
\ea
Then the time-evolution of the primary constraints generates the secondary constraints 
\ba
\calC^{(2)}_{\Bhi}  &\equiv& \{ \calC^{(2)}_{\Bhi} \,, \calH^V_{T} \}  
= 4\mu_{h1} \Bhi +2n_1 \Bfi - k \piFhi \approx0 \,,\\
\calC^{(2)}_{\Bfi}  &\equiv& \{ \calC^{(2)}_{\Bfi} \,, \calH^V_{T} \}  
= 2n_1 \Bhi +4\mu_{f1} \Bfi  - k \piFfi \approx0 \,,
\ea
and the time-evolution of the secondary constraints gives
\ba
\begin{pmatrix}
	{\dot \calC}^{(2)}_{\Bhi}  \\
	{\dot \calC}^{(2)}_{\Bfi} 
\end{pmatrix}
=
\begin{pmatrix}
	\{ \calC^{(2)}_{\Bhi} \,, \calH^V_{T} \}    \\
	\{ \calC^{(2)}_{\Bfi} \,, \calH^V_{T} \}  
\end{pmatrix}
=
\begin{pmatrix}
	\{ \calC^{(2)}_{\Bhi} \,, \calH^V \}    \\
	\{ \calC^{(2)}_{\Bfi} \,, \calH^V \}  
\end{pmatrix}
+
\begin{pmatrix}
	4\mu_{h1}&2n_1\\
	2n_1 &4\mu_{f1}
\end{pmatrix}
\begin{pmatrix}
	\lambda_{\Bhi}  \\
	\lambda_{\Bfi} 
\end{pmatrix}
 \approx0 \,.
 \ea
Therefore, when $n_1^2 - 4\mu_{f1}\mu_{h1}\neq 0$, namely the coefficient matrix in front of the Lagrange multipliers is not degenerate, all the Lagrange multipliers $\lambda_{\Bhi}$ and $\lambda_{\Bfi}$ are determined by the above equations, and all the primary and secondary constraints are second class. 
In this case, the total number of physical DOFs is $(8 \times 2 - 8)/2=4$, and thus we disregard this option. 

On the other hand, when $n_1^2 - 4\mu_{f1}\mu_{h1}= 0$, the coefficient matrix in front of the Lagrange multipliers is degenerate, and two out of four Lagrange multipliers cannot be determined. Hereafter we assume $\mu_{h1} \neq 0$ and solve $n_1^2 - 4\mu_{f1}\mu_{h1}= 0$ for $\mu_{f1}$\footnote{For $\mu_{h1} =0$ case, one can simply switch all the notation of $h$ and $f$. When $\mu_{h1}=\mu_{f1}=n_1=0$, all the constraints becomes first-class, implying that the physical degrees of freedom is zero in the vector sector.}. It is convenient to redefine the primary constraints associated with ${\Bfi}$ as a linear combination of the original primary constraints : 
\ba
\calCt^{(1)}_{\Bhi} &\equiv& \pi_{\Bhi}  \approx 0 \,,\\
\calCt^{(1)}_{\Bfi} &\equiv& \pi_{\Bfi} - \frac{n_1}{2\mu_{h1}} \piBhi \approx 0 \,.
\ea
We also redefine the total Hamiltonian 
\ba
\calH^V_{T} = \calH^V 
+ \lambdat_{\Bhi} \calCt^{(1)}_{\Bhi} 
+ \lambdat_{\Bfi} \calCt^{(1)}_{\Bfi} \,.
\ea
Then the secondary constraints become 
\ba
\calCt^{(2)}_{\Bhi}  &\equiv& \{ \calCt^{(1)}_{\Bhi} \,, \calH^V_{T} \}  
= 4\mu_{h1} \Bhi +2n_1 \Bfi - k \piFhi\approx0 \,,\\
\calCt^{(2)}_{\Bfi}  &\equiv& \{ \calCt^{(1)}_{\Bfi} \,, \calH^V_{T} \}  
=k\left( \frac{n_1}{2\mu_{h1}} {\piFhi}- \piFfi\right)\approx0 \,.
\ea
The time-evolution of the secondary constraints yields
\ba
&&{\dot \calCt}{}^{(2)}_{\Bhi} = \{ \calCt^{(2)}_{\Bhi} \,, \calH^V_{T} \}  
={2k\left(2\mu_{h1}\Fhi+n_1\Ffi\right)}
+ 4\mu_{h1}
\lambdat_{\Bhi} 
   \approx0 
\,,\\
&&{\dot \calCt}{}^{(2)}_{\Bfi} =\{ \calCt^{(2)}_{\Bfi} \,, \calH^V_{T} \}\approx0 
\,.
\ea
Here, the time-evolution of the secondary constraints $\calCt^{(2)}_{\Bfi}$ is trivially zero. Therefore, two of the Lagrange multipliers can be determined by the time evolution of the secondary constraints, and the rest of them are undetermined. Since the constraints $\calCt^{(1)}_{\Bfi}$ and $\calCt^{(2)}_{\Bfi}$ commute with all the constraints including themselves, 
they are first-class constraints. 
To summarize, we find
\begin{eqnarray}
	\textrm{Vector DOF} = \frac{8\times 2 - 4 \,(\textrm{2 primary \& 2 secondary )} - 4 \,(\textrm{2 primary \& 2 secondary})  \times 2 \,(\textrm{first-class})}{2} 
	=2 \,.
\end{eqnarray}
The choice of the coefficients are 
\ba
\kappa_{h2}=-2\kappa_{h1} \neq 0 \,,\qquad
\kappa_{f2}=-2\kappa_{f1}\neq 0 \,, \qquad
l_2=0 \,, \qquad
\mu_{f1}=\frac{n_1^2}{4\mu_{h1}} \,, \qquad
\mu_{h1} \neq 0 \,.
\label{ConditionV}
\ea
In the analysis for the scalar sector in the next section, the conditions \eqref{ConditionT} and \eqref{ConditionV} are imposed.

\section{Scalar sector}
In this section, we focus on the scalar sector. Here, we need to eliminate $7$ DOFs in the scalar sector in order to have $1$ physical DOF. Introducing dimensionless variables, $\beta_\Phi \to\beta_\Phi/k$ and ${\cal E}_\Phi\to{\cal E}_\Phi/k^2$, the Lagrangian reduces to
\ba
	\calL^S 
	= \calL^S_{\rm kin} + \calL^S_{\rm cross} + \calL^S_{\rm mass} 
	\,,
\ea
 where the explicit form of the first part reads
\ba
{\cal L}_{\mathrm{kin}}^S &=&
4 (\kappa_{h1}+\kappa_{h2}+\kappa_{h3}+\kappa_{h4}) \dot{\alpha}_h^2
- (2 \kappa_{h1}+\kappa_{h2}) \dot\beta_h^2 
+ 12 (\kappa_{h1}+3 \kappa_{h4}) \dot{{\cal R}}_h^2
+ 4 (\kappa_{h1}+\kappa_{h4}) \dot{{\cal E}}_h^2 
\notag\\
&& 
- 4 (\kappa_{h3}+2 \kappa_{h4})\left( -3\dot{{\cal R}}_h + \dot{{\cal E}}_h 
\right) \dot{\alpha}_h
-8 (\kappa_{h1}+3 \kappa_{h4}) \dot{{\cal R}}_h \dot{{\cal E}}_h 
\notag\\
&& 
+ 4 (\kappa_{f1}+\kappa_{f2}+\kappa_{f3}+\kappa_{f4}) \dot{\alpha}_f^2
- (2 \kappa_{f1}+\kappa_{f2}) \dot\beta_f^2 
+ 12 (\kappa_{f1}+3 \kappa_{f4}) \dot{{\cal R}}_f^2
+ 4 (\kappa_{f1}+\kappa_{f4}) \dot{{\cal E}}_f^2 
\notag\\
&& 
- 4 (\kappa_{f3}+2 \kappa_{f4})\left( -3\dot{{\cal R}}_f + \dot{{\cal E}}_f 
\right) \dot{\alpha}_f
-8 (\kappa_{f1}+3 \kappa_{f4}) \dot{{\cal R}}_f \dot{{\cal E}}_f 
\notag\\
&&
+ 4 \left( l_2+l_3+l_4+l_5\right)\dot\alpha_h\dot\alpha_f-l_2\dot\beta_h\dot\beta_f+36l_4\dot{\cal R}_h\dot{\cal R}_f+4l_4\dot{\cal E}_h\dot{\cal E}_f
+12\left( l_3+l_4\right)\dot\alpha_h\dot{\cal R}_f+12\left( l_4+l_5\right)\dot\alpha_f\dot{\cal R}_h
\notag\\
&&-4\left( l_3+l_4\right)\dot\alpha_h\dot{\cal E}_f-4\left( l_4+l_5\right)\dot\alpha_f\dot{\cal E}_h-12l_4\left(\dot{\cal R}_h\dot{\cal E}_f+\dot{\cal R}_f\dot{\cal E}_h\right)
\,.
\ea
 and the explicit expression for the remaining parts can be found in the Appendix \ref{app:Lscalar}.
Once we impose the condition \eqref{ConditionV}, the time derivative of $\beta_h$ and $\beta_f$ vanishes in the Lagrangian,
hence $\beta_h\,,\beta_f$ can be treated as the non-dynamical variables.
By utilizing the field redefinition summarized in Appendix \ref{sec:field redef}, we can further impose without loss of generality
\ba
\kappa_{h3}=2\kappa_{h1}, \qquad \kappa_{f3}=2\kappa_{f1}, \qquad l_3+l_4+l_5=0 \,,
\label{FRDV}
\ea
in addition to $l_1=0$. Hereafter, we assume these conditions to simplify the discussion.
The conjugate momenta can be written as
\ba
	\begin{pmatrix}
		\pi_{\alpha_h}\\
		\pi_{{\cal R}_h}\\
		\pi_{{\cal E}_h}\\
		\pi_{\alpha_f}\\
		\pi_{{\cal R}_f}\\
		\pi_{{\cal E}_f}\\
	\end{pmatrix}
		=4\calK_S
	\begin{pmatrix}
		\dot\alpha_h\\
		\dot{\cal R}_h\\
		\dot{\cal E}_h\\
		\dot\alpha_f\\
		\dot{\cal R}_f\\
		\dot{\cal E}_f\\
	\end{pmatrix}+
	4
	\begin{pmatrix}
		0\\
		-4\kappa_{h1}\\
		0\\
		-l_3\\
		-3l_3\\
		l_3\\
	\end{pmatrix}
	k\beta_h
	+4
	\begin{pmatrix}
		-l_5\\
		-3l_5\\
		l_5\\
		0\\
		-4\kappa_{f1}\\
		0\\
	\end{pmatrix}
	k\beta_f \,,
\ea
where the kinetic matrix for the scalar variables $\{\alpha_h,{\cal R}_h,{\cal E}_h\,,\alpha_f,{\cal R}_f,{\cal E}_f\}$ is given by
\ba
	{\cal K}_S=
	\begin{pmatrix}
		2(\kappa_{h1}+\kappa_{h4})& 6(\kappa_{h1}+\kappa_{h4})&
		-2(\kappa_{h1}+\kappa_{h4})& 
		0& 
		-3l_5& 
		l_5 \\
		* &
		6(\kappa_{h1}+3\kappa_{h4}) & -2(\kappa_{h1}+3\kappa_{h4}) & 
		-3l_3 & 
		-9(l_3+l_5) & 
		3(l_3 + l_5)\\
		* &*&
		2(\kappa_{h1}+\kappa_{h4})&
		l_3&
		3(l_3+l_5)&
		-(l_3+l_5)\\
		*&*&*&
		2(\kappa_{f1}+\kappa_{f4})&
		6(\kappa_{f1}+\kappa_{f4})&
		-2(\kappa_{f1}+\kappa_{f4})\\
		*&*&*&*&
		6(\kappa_{f1}+3\kappa_{f4})&
		-2(\kappa_{f1}+3\kappa_{f4})\\
		*&*&*&*&*&2(\kappa_{f1}+\kappa_{f4})\\
	\end{pmatrix}
	\,,
\ea
and
\ba
	&&\pi_{\beta_h}=\pi_{\beta_f}=0
	\label{momentumb}
	\,.
\ea
Note that there are at least two primary constraints from $\pi_{\beta_h} {=0}$ and $\pi_{\beta_f} {=0}$.
The Hamiltonian is given by
\ba
\calH^S&=&\dot\alpha_h\pi_{\alpha_h}+\dot\alpha_f\pi_{\alpha_f}+\dot{\cal R}_h\pi_{{\cal R}_h}+\dot{\cal R}_f\pi_{{\cal R}_f}+\dot{\cal E}_h\pi_{{\cal E}_h}+\dot{\cal E}_f\pi_{{\cal E}_f}
+\dot\beta_h\pi_{\beta_h}+\dot\beta_f\pi_{\beta_f}-\calL^S\,.
\ea

\subsection{Classification of primary constraints}
Now, we would like to classify the cases based on the number of primary constraints. 
As performed in the analysis of the vector sector, we consider the eigenvalue equation, 
\ba
{\cal F}_S(\lambda) \equiv {\rm det} ({\cal K}_S - \lambda I) \,,
\ea
The eigenvalue equation with $\lambda = 0$, namely the determinant of the kinetic matrix, reads
\ba
	{\rm det} \, {\cal K}_S={\cal F}_S(0) =16\kappa_{h1}\kappa_{f1}
			\Bigl[ 8(\kappa_{f1}+\kappa_{f4})\kappa_{h1}+3l_3^2\Bigr]\Bigl[ 8(\kappa_{h1}+\kappa_{h4})\kappa_{f1}+3l_5^2\Bigr]
	\,.
\ea
When the above determinant is non-zero, that is, $ \det{\cal K}_S\neq 0$, there are only two primary constraints, which can be defined by \eqref{momentumb}. In this case, the number of the physical DOFs is four as proved in Appendix~\ref{app:twoprimary}. 
Therefore, we disregard this option. 
The case of $3$ primary constraint can be obtained by demanding ${\cal F}_S(0) =0 $, 
\ba
&& \textrm{3 primary constraints : }\quad {\cal F}_S(0) =0 \quad \longleftrightarrow\quad 
	\kappa_{f4}=-\kappa_{f1}-\frac{3l_3^2}{8\kappa_{h1}} \qquad
	{\rm or}\qquad
	\kappa_{h4}=-\kappa_{h1}-\frac{3l_5^2}{8\kappa_{f1}}
	\,.
		\label{S3Pb}
\ea
Using the conditions above, we have 
\ba
{\cal F}_S'(0) = 
  \begin{cases}
	\, -8 \kappa_{f1} (5 l_3^2 + 8\kappa_{h1}^2) (3l_5^2 + 8 \kappa_{f1} (\kappa_{h1} + \kappa_{h4}))\quad {\rm for} \quad \displaystyle{\kappa_{f4}=-\kappa_{f1}-\frac{3l_3^2}{8\kappa_{h1}}}  \,, \\\\
	\, -8 \kappa_{h1} (5l_5^2 + 8\kappa_{f1}^2) (3 l_3^2 + 8\kappa_{h1}(\kappa_{f1} + \kappa_{f4}))\quad {\rm for} \quad \displaystyle{\kappa_{h4}=-\kappa_{h1}-\frac{3l_5^2}{8\kappa_{f1}}}  \,.
\end{cases}
\label{fsd}
\ea
Now, ${\cal F}_S'(0) = 0$ gives only one solution of $4$ primary case due to the symmetric property under $h$ and $f$ in \eqref{S3Pb}, 
\ba
&& \textrm{4 primary constraints : }\quad {\cal F}_S(0) =0 \quad \& \quad {\cal F}_S'(0) =0 \quad
\longleftrightarrow\quad
\kappa_{h4}=-\kappa_{h1}-\frac{3l_5^2}{8\kappa_{f1}} \quad \& \quad
\kappa_{f4}=-\kappa_{f1}-\frac{3l_3^2}{8\kappa_{h1}}\,.
\label{S4P}
\ea
The absence of the case with five primary constraints can be proved as follows. 
In addition to ${\cal F}_S(0) =0$ and ${\cal F}_S'(0) =0$, we further need to impose ${\cal F}_S''(0) =0$, which is given by
\ba
{\cal F}_S''(0)  = 32 (l_5^2 + 16 \kappa_{f1}^2) 
\left(\kappa_{h1} - \frac{9l_3l_5\kappa_{f1}}{l_5^2 + 16 \kappa_{f1}^2}\right)^2 
+\frac{8l_3^2(5l_5^2+8\kappa_{f1}^2)^2}{l_5^2+16\kappa_{f1}^2}
=0 \,.
\ea
It is manifest that there is no real solution for this equation under the assumption \eqref{ConditionT}, and therefore, the scalar sector cannot have 5 or more primary constraints.

\subsection{3 primary constraints}
In this subsection, we consider the case with $3$ primary constraints.
Although there are two options as in \eqref{S3Pb}, they are essentially equivalent since they are transformed each other as shown in Appendix~\ref{app:3primaryredef}, which satisfies :
\ba
&&\kappa_{h2}= -\kappa_{h3} = -2\kappa_{h1} \neq 0 \,, \qquad
\kappa_{f2}= -\kappa_{f3} =-2\kappa_{f1} \neq 0 \,, \qquad \kappa_{f4} = -\kappa_{f1} - \frac{3l_3^2}{8\kappa_{h1}}\,,\nonumber\\
&&\
l_1 = l_2=l_4=\mu_{f1}=n_1=0 \,, \qquad l_5=-l_3 \,.
\label{FRDS3}
\ea
In this case, we have three primary constraints, which are defined by :
\ba
	\calC^{(1)}_{\alpha_f}\equiv\pi_{\alpha_f} 
	- {l_3 \over 4\kappa_{h1}}(\pi_{\calR_h} - 3 \pi_{\alpha_h}) \approx 0
	\,,\qquad \calC^{(1)}_{\beta_h}\equiv\pi_{\beta_h} \approx 0
	\,,\qquad 
	\calC^{(1)}_{\beta_f}\equiv\pi_{\beta_f} \approx 0
	\,.
\ea
The total Hamiltonian is given by
\ba
	\calH_T^S=\calH^S+\lambda_{\alpha_f}\calC^{(1)}_{\alpha_f}+\lambda_{\beta_h}\calC^{(1)}_{\beta_h}+\lambda_{\beta_f}\calC^{(1)}_{\beta_f}
	\,.
\ea
The evolution of the primary constraints yields the secondary constraints
\ba
	&& \calC^{(2)}_{\alpha_f}
		\equiv\{\calC^{(1)}_{\alpha_f},\calH_T^S\}
        {=- 2 \left( 2k^2 l_3 + 2 n_2 + {3 l_3 \mu_{h1} \over \kappa_{h1}}\right)\alpha_h
		+2 \left(2k^2 l_3 -6 n_2 + {3 l_3 \mu_{h1} \over \kappa_{h1}}\right) \calR_h
		+ 2 \left(2k^2 l_3 + 2 n_2 - {l_3 \mu_{h1} \over \kappa_{h1}}\right)\calE_h
		}	
		 \nonumber\\
		&&\qquad\qquad\qquad\qquad~~~
		+ 4 \left({k^2 l_3^2 \over \kappa_{h1}}-2\mu_{f2}\right)(\alpha_f-\calE_f)
		+ 4 \left( {k^2 (3l_3^2 + 4\kappa_{f1} \kappa_{h1})\over \kappa_{h1}} 
		- 6\mu_{f2} \right) \calR_f \approx 0 \,, \\
	&& \calC^{(2)}_{\beta_h}
		\equiv\{\calC^{(1)}_{\beta_h},\calH_T^S\}
		=-k\left(\pi_{\alpha_h}+\pi_{{\cal E}_h}\right) +4\mu_{h1}\beta_h \approx 0	\,,\\
	&& \calC^{(2)}_{\beta_f}
		\equiv\{\calC^{(1)}_{\beta_f},\calH_T^S\}
		=-k \pi_{\calE_f} 
		+\frac{k l_3}{4\kappa_{h1}}(3\pi_{\alpha_h}-\pi_{\calR_h}) \approx 0 \,.
\ea
Here, none of the above constraints can be trivially zero with any choice of the parameters under the assumption \eqref{ConditionT}.
Then, the time evolution of the secondary constraint $\calC^{(2)}_{\beta_f}$ gives the tertiary constraint
\ba
	&&\calC^{(3)}_{\beta_f}\equiv\{\calC^{(2)}_{\beta_f},\calH_T^S\} 
= k \,\calC^{(2)}_{\alpha_f} + 4 k^3 
\Big[
l_3 (\alpha_h -\calE_h + 3\calR_h)-4 \kappa_{f1} \calR_f
\Big] \approx 0
\,,
\ea
and its time evolution demands
\ba
{\dot \calC}^{(3)}_{\beta_f} = k ({\dot \calC}^{(2)}_{\alpha_f} + k\calC^{(2)}_{\beta_f}) \approx 0 \,.
\ea
Since ${\dot \calC}^{(3)}_{\beta_f} $ does not generate an independent equation, there is no more constraint from $\beta_f$. The evolution of the rest of the secondary constraints are given by
\ba
\begin{pmatrix}
	{\dot \calC}^{(2)}_{\alpha_{f}}  \\
	{\dot \calC}^{(2)}_{\beta_{h}} 
\end{pmatrix}
=
\begin{pmatrix}
	\{ \calC^{(2)}_{\alpha_{f}} \,, {\calH^S_{T}} \}    \\
	\{ \calC^{(2)}_{\beta_{h}} \,, {\calH^S_{T}} \}  
\end{pmatrix}
=
\begin{pmatrix}
	\{ \calC^{(2)}_{\alpha_{f}} \,, {\calH^S} \}    \\
	\{ \calC^{(2)}_{\beta_{h}} \,, {\calH^S} \}  
\end{pmatrix}
+
 2
\begin{pmatrix}
	\displaystyle{ -4 \mu_{f2} - {3 l_3^2 \mu_{h1} \over \kappa_{h1}^2}}&0\\
	0&2\mu_{h1}	
\end{pmatrix}
\begin{pmatrix}
	\lambda_{\alpha_{f}}  \\
	\lambda_{\beta_{h}} 
\end{pmatrix}
\approx 0 \,.
\ea
Since $\mu_{h1} \neq 0$, the Lagrange multiplier $\lambda_{\beta_h}$ is determined by ${\dot \calC}^{(2)}_{\beta_h}=0$.
When the Poisson bracket $\{\calC^{(2)}_{\alpha_f},\calC^{(1)}_{\alpha_f}\}$, i.e., the coefficient of $\lambda_{\alpha_f}$ in ${\dot \calC}^{(2)}_{\alpha_{f}}$, is non-vanishing, the evolution of $\calC^{(2)}_{\alpha_f}$ determines the Lagrange multiplier $\lambda_{\alpha_f}$
and no more constraint from $\calC^{(2)}_{\alpha_f}$ will be generated. 
Now we redefine the following constraints : 
\ba
\widetilde{\calC}^{(2)}_{\beta_f} \equiv \calC^{(2)}_{\beta_f} - k \calC^{(1)}_{\alpha_f} \approx 0 \,, \qquad 
\widetilde{\calC}^{(3)}_{\beta_f} \equiv \calC^{(3)}_{\beta_f} - k \calC^{(2)}_{\alpha_f} \approx 0 \,. 
\ea
The constraints, $\calC^{(1)}_{\beta_f} $ and 
$\widetilde{\calC}^{(2,3)}_{\beta_f} $ commute with all constraints, therefore, these are first class. The rest of constraints are second class. In summary, the number of the physical DOFs is $(8\times 2 - 4 -3\times 2)/2 = 3$. 

In order to eliminate extra DOFs, one has to impose an extra condition $\{\calC^{(2)}_{\alpha_f}\,,\calC^{(1)}_{\alpha_f}\} =0$, namely
\ba
	\mu_{f2} = -{3 l_3^2 \mu_{h1} \over 4 \kappa_{h1}^2}
	\,.
	\label{Con:S3Pb:tertiary}
\ea
In this case, the evolution of $\calC^{(2)}_{\alpha_f}$ yields the tertiary constraint when $l_3 \neq 0$ or $n_2 \neq 0$,
\ba
	\calC^{(3)}_{\alpha_f}&\equiv&\{\calC^{(2)}_{\alpha_f},\calH_T^S\} \approx 0
			\,,
\label{Caf3}
\ea
where the explicit expression of $\calC^{(3)}_{\alpha_f}$ is given in Appendix \ref{app:constraints}.
When $l_3 = n_2 =0$, $\calC^{(3)}_{\alpha_f}= -k \calC^{(2)}_{\beta_f}$, implying that there is no more constraint. Therefore, we, hereafter, consider the case with   $l_3 \neq 0$ or $n_2 \neq 0$. 
Now, the evolution of $\calC^{(3)}_{\alpha_f}$ yields the 
quaternary constraint $\calC^{(4)}_{\alpha_f}=\{\calC^{(3)}_{\alpha_f},\calH^S_T\} \approx 0$.
Since $\calC^{(4)}_{\alpha_f}$ contains $\lambda_{\beta_h}$, it is useful to define the following 
 linear combination of constraints :
\ba
\widetilde{\calC}^{(3)}_{\alpha_f} &\equiv& {\calC}^{(3)}_{\alpha_f} + {2 k l_3 \over \kappa_{h1}} {\calC}^{(2)}_{\beta_h}{+k\calC^{(2)}_{\beta_f}} \approx 0 \,, 
\label{Caf34b} \\
\widetilde\calC^{(4)}_{\alpha_f} &\equiv& \{\widetilde\calC^{(3)}_{\alpha_f},\calH^S_T\} \approx 0 \,.
\label{C4a}
\ea
When $\{{\widetilde\calC^{(4)}_{\alpha_f}},\calC^{(1)}_{\alpha_f}\}\neq 0$, 
the Lagrange multiplier $\lambda_{\alpha_f}$ is determined by the time evolution of ${\widetilde\calC^{(4)}_{\alpha_f}}$.
In this case, the constraints, $\calC^{(1)}_{\beta_f} $ and 
$\widetilde{\calC}^{(2,3)}_{\beta_f} $ still commute with all constraints, and hence, these are first class. The rest of constraints are second class, therefore, the number of the physical DOFs is 
given by $(8\times 2 -6 -3\times 2)/2 = 2$. 

To obtain a theory with $1$ DOF 
 in the scalar sector, one more DOF has to be eliminated. Then, we would like to consider the following case
\ba
\{{\widetilde\calC^{(4)}_{\alpha_f}},\calC^{(1)}_{\alpha_f}\}
={\frac{4}{\kappa_{h1}^2}\left(-\frac{3l_3^2\mu_{h1}^2}{\kappa_{h1}}+\frac{\kappa_{f1}(2\kappa_{h1}n_2+3l_3\mu_{h1})^2}{8\kappa_{f1}(\kappa_{h1}+\kappa_{h4})+3l_3^2}\right)}
= 0 \,.
\ea 
Solving the above equation, we obtain
\ba
\kappa_{h4}
= {-\kappa_{h1}-\frac{3l_3^2}{8\kappa_{f1}}+\frac{\kappa_{h1}(2\kappa_{h1}n_2+3l_3\mu_{h1})^2}{24l_3^2\mu_{h1}^2}}
\,.
\label{LCon:3primary}
\ea
In this case, we have two additional constraints : 
\ba
\widetilde\calC^{(5)}_{\alpha_f}\equiv\{\widetilde\calC^{(4)}_{\alpha_f},\calH^S_T\} \approx 0 \,, \qquad 
\widetilde\calC^{(6)}_{\alpha_f}\equiv\{\widetilde\calC^{(5)}_{\alpha_f},\calH^S_T\} \approx 0 \,.
\label{Caf56b}
\ea
Again, $\widetilde\calC^{(6)}_{\alpha_f}$ contains the Lagrange multiplier $\lambda_{\beta_h}$, we redefine 
the constraint as 
\ba
\overline{\calC}{}^{(5)}_{\alpha_f} \equiv {\widetilde\calC}^{(5)}_{\alpha_f} + x \,  {\calC}^{(2)}_{\beta_h} \approx 0 \,, \qquad 
\overline\calC{}^{(6)}_{\alpha_f}\equiv\{\widetilde\calC^{(5)}_{\alpha_f},\calH^S_T\} \approx 0 \,,
\ea
where 
\ba
x = 
\frac{k l_3 }{ \kappa_{h1}^3}
\biggl[
{-2}k^2 \kappa_{h1}^2 + \frac{\mu_{h1}(8n_2 \kappa_{f1} \kappa_{h1}^2 -9 l_3^3 \mu_{h1})}{\kappa_{f1}(2\kappa_{h1}n_2  + 3 l_3 \mu_{h1})}
\biggr] \,.
\ea
If $\{\overline\calC{}^{(6)}_{\alpha_f},\calC^{(1)}_{\alpha_f}\}\neq 0$, the Lagrange multiplier $\lambda_{\alpha_f}$ can be determined by $\dot{ \overline\calC}{}^{(6)}_{\alpha_f}=\{\overline\calC^{(6)}_{\alpha_f},\calH^S_T\} \approx 0 $ and no further constraint is generated.
\ba
 \rm{first \, class} \quad &:& \quad \calC^{(1)}_{\beta_f} \,, \qquad \widetilde\calC^{(2)}_{\beta_f} \,, \qquad \widetilde\calC^{(3)}_{\beta_f} \,, \\
 \rm{second \, class} \quad &:& \quad \calC^{(1)}_{\alpha_f} \,, \qquad \calC^{(2)}_{\alpha_f} \,, \qquad \widetilde\calC^{(3)}_{\alpha_f} \,, \qquad \widetilde\calC^{(4)}_{\alpha_f} \,, \qquad \overline\calC^{(5)}_{\alpha_f} \,, \qquad \overline\calC^{(6)}_{\alpha_f} \,, \qquad \calC^{(1)}_{\beta_h} \,, \qquad \calC^{(2)}_{\beta_h} \,,
\ea
we finally have
\ba
\textrm{Scalar DOF} &=& \frac{1}{2}\times
\Bigl[
8\times 2 - 8 \,(\textrm{2 primary \& 2 secondary \& 1 tertiary \& 1 quaternary + 2 more)} \notag\\
&&~~~~~~~
- 3 \,(\textrm{1 primary \& 1 secondary \& 1 tertiary)} \times 2 \,(\textrm{first-class})
\Bigr] 
=1 \,.
\ea
To summarize, we find a novel class of theory, \\
{\bf  (Class~Ia)} :
\ba
&&\kappa_{h2}= -\kappa_{h3} = -2\kappa_{h1} \neq 0 \,, \qquad
\kappa_{f2}= -\kappa_{f3}= -2\kappa_{f1} \neq 0 \,, \qquad 
\kappa_{f4} = - \kappa_{f1} - {3l_3^2 \over 8\kappa_{h1}} \,, \qquad
{\mu_{f2}=-\frac{3l_3^2\mu_{h1}}{4\kappa_{h1}^2}} \,,
\nonumber\\
&&\
l_1 = l_2 = l_4= \mu_{f1}=n_1=0 \,, \qquad 
l_5 = -l_3 \,, \qquad 
\kappa_{h4}= {-\kappa_{h1}-\frac{3l_3^2}{8\kappa_{f1}}+\frac{\kappa_{h1}(2\kappa_{h1}n_2+3l_3\mu_{h1})^2}{24l_3^2\mu_{h1}^2} \,.}
\label{ClassI}
\ea
The Lagrangian for class with 1 DOF in the scalar sector Ia is given by
\ba
{\cal L} &=&
-\left( \kappa_{h1} h_{\mu\nu}\widehat{\cal E}^{\mu\nu\alpha\beta}h_{\alpha\beta}+\kappa_{f1}f_{\mu\nu}\widehat{\cal E}^{\mu\nu\alpha\beta}f_{\alpha\beta}\right)
+\delta\kappa_{h4} h_{,\mu} f^{,\mu}
+{3l_3^2 \over 8 \kappa_{h1}} f_{,\mu} f^{,\mu}
+ l_3 \left(h_{,\nu} f^{\mu\nu}_{~~,\mu} -
h^{\mu\nu}_{~~,\mu} f_{,\nu} 
\right)\nonumber\\
&&
-\mu_{h1} h_{\mu\nu}h^{\mu\nu}
-\mu_{h2} h^2
+\left({{3l_3^2\mu_{h1} \over 4 \kappa_{h1}^2}} f - n_2 h \right)f \,,
\ea
where 
\ba
\delta\kappa_{h4} = \frac{3l_3^2}{8\kappa_{f1}}-\frac{\kappa_{h1}(2\kappa_{h1}n_2+3l_3\mu_{h1})^2}{24l_3^2\mu_{h1}^2}\,.
\ea
One can check that this theory is invariant under the gauge transformation
\ba
h_{\mu\nu} &\to& {\widetilde h}_{\mu\nu} = 
h_{\mu\nu} \,,\\
f_{\mu\nu} &\to& {\widetilde f}_{\mu\nu} =
f_{\mu\nu} +\partial_\mu \xi_\nu+  \partial_\nu \xi_\mu \qquad {\rm with}\qquad\partial^\mu \xi_\mu=0 \,.
\label{TDiff}
\ea
As one can see from the transverse condition in the gauge transformation, this class is totally distinct from the linearized Hassan-Rosen bigravity, and there are non-trivial kinetic terms for $h$, derivative and mass interactions. 
\\

Let us finally discuss the final option where the time-evolution of the tertiary constraint 
$\dot{\widetilde\calC}{}^{(3)}_{\alpha_f}$ does not yield new constraint. Such a case can be found by rewriting $\dot{\widetilde\calC}{}^{(3)}_{\alpha_f}$ in terms of other constraints, $\calC^{(1)}_{\beta_f}$, $\widetilde\calC^{(2)}_{\beta_f}$, $\widetilde\calC^{(3)}_{\beta_f} $, $\calC^{(1)}_{\alpha_f} $, $ \calC^{(2)}_{\alpha_f}$, $\calC^{(1)}_{\beta_h} $, and $\calC^{(2)}_{\beta_h} $,  and setting it to be zero. Then, we obtain two conditions : Eq.~\eqref{LCon:3primary} and 
\ba
\frac{24 l_3^2 \mu_{h1}^3 (4n_2^2 \kappa_{h1}^2 + 3l_3^2 \mu_{h1} (\mu_{h1} + 4\mu_{h2}))}{\kappa_{h1}^4 (2n_2 \kappa_{h1} + 3l_3 \mu_{h1})^2} =0\,,
\ea
which can be solved for $\mu_{h2}$,
\ba
\mu_{h2} = -\frac{n_2^2 \kappa_{h1}^2}{3 l_3^2 \mu_{h1}} -{\mu_{h1} \over 4} \,,
\label{classIb}
\ea
since we assumed $l_3 \neq 0$ and $\mu_{h1}\neq 0$. Note that the case where
$\{\overline\calC{}^{(6)}_{\alpha_f},\calC^{(1)}_{\alpha_f}\} $ vanishes in the Class Ia reduces to this option.  
In this case, as shown in \eqref{eq:C^4_alpha_f}, the time-evolution of both the tertiary constraints $\widetilde\calC^{(3)}_{\alpha_f}$ and $\widetilde\calC^{(3)}_{\beta_f}$ can be written in terms of the linear combination of the primary and second class constraints, implying no further constraints. Redefining $ \calC^{(2)}_{\alpha_f}$ as 
\ba
  \widetilde\calC^{(2)}_{\alpha_f} \equiv  \calC^{(2)}_{\alpha_f} + {2k l_3 \over \kappa_{h1}} \calC^{(1)}_{\beta_h} \,,
\ea 
we find 
\ba
\rm{first \, class} \quad &:& \quad \calC^{(1)}_{\alpha_f} \,, \qquad \widetilde\calC^{(2)}_{\alpha_f} \,, \qquad \widetilde\calC^{(3)}_{\alpha_f} \,, \qquad \calC^{(1)}_{\beta_f} \,, \qquad \widetilde\calC^{(2)}_{\beta_f} \,, \qquad \widetilde\calC^{(3)}_{\beta_f} \,, \\
\rm{second \, class} \quad &:& \quad  \calC^{(1)}_{\beta_h} \,, \qquad \calC^{(2)}_{\beta_h} \,,
\ea
and 
\ba
\textrm{Scalar DOF} &=& \frac{1}{2}\times
\Bigl[
8\times 2 - 2 \,(\textrm{1 primary \& 1 secondary)} \notag\\
&&~~~~~~~
- 6 \,(\textrm{2 primary \& 2 secondary \& 2 tertiary)} \times 2 \,(\textrm{first-class})
\Bigr] 
=1 \,.
\ea
In this case, the number of the physical DOF is the same as the Class Ia, and the resultant theory is invariant under the gauge transformation \eqref{TDiff}.
For this reason, this case can be considered as the special case of the Class I by choosing \eqref{classIb} although an additional gauge symmetry is present.  
To summarize, 
\\
{\bf  (Class~Ib)} :
\ba
&&\kappa_{h2}= -\kappa_{h3} = -2\kappa_{h1} \neq 0 \,, \qquad
\kappa_{f2}= -\kappa_{f3}= -2\kappa_{f1} \neq 0 \,, \qquad 
\kappa_{f4} = - \kappa_{f1} - {3l_3^2 \over 8\kappa_{h1}} \,, \qquad
{\mu_{f2}=-\frac{3l_3^2\mu_{h1}}{4\kappa_{h1}^2}} \,,
\nonumber\\
&&\
l_1 = l_2 = l_4= \mu_{f1}=n_1=0 \,, \qquad 
l_5 = -l_3 \,, \qquad 
\kappa_{h4}= -\kappa_{h1}-\frac{3l_3^2}{8\kappa_{f1}}+\frac{\kappa_{h1}(2\kappa_{h1}n_2+3l_3\mu_{h1})^2}{24l_3^2\mu_{h1}^2}\,, \nonumber \\
&&
\mu_2 = -\frac{n_2^2 \kappa_{h1}^2}{3 l_3^2 \mu_{h1}} -{\mu_{h1} \over 4} \,.
\ea
The Lagrangian for 
the Class Ib is given by
\ba
{\cal L} &=&
-\left( \kappa_{h1} h_{\mu\nu}\widehat{\cal E}^{\mu\nu\alpha\beta}h_{\alpha\beta}+\kappa_{f1}f_{\mu\nu}\widehat{\cal E}^{\mu\nu\alpha\beta}f_{\alpha\beta}\right)
+\delta\kappa_{h4} h_{,\mu} f^{,\mu}
+{3l_3^2 \over 8 \kappa_{h1}} f_{,\mu} f^{,\mu}
+ l_3 \left(h_{,\nu} f^{\mu\nu}_{~~,\mu} -
h^{\mu\nu}_{~~,\mu} f_{,\nu} 
\right)\nonumber\\
&&
{-\mu_{h1}\left( h_{\mu\nu}h^{\mu\nu}+\frac{1}{4}h^2\right)+\frac{1}{3\mu_{h1}}\left(\frac{\kappa_{h1}n_2}{l_3}h-\frac{3l_3\mu_{h1}}{2\kappa_{h1}}f\right)^2}
\,.
\ea

\subsection{$4$ primary constraints}
Next, let us consider the case with four primary constraints.
As you can see from \eqref{S4P}, there are kinetic interactions between $h$ and $f$ fields, which will make the Hamiltonian analysis involved in general.
It is interesting to note that we can always map this theory into a simpler theory with two Einstein-Hilbert terms without kinetic interactions between them as explicitly shown in \ref{app:4primaryredef}.
Hereinafter we will perform the Hamiltonian analysis in this simple model :
\ba
&&\kappa_{h2}= -\kappa_{h3} = 2\kappa_{h4}= -2\kappa_{h1} \neq 0 \,, \qquad
\kappa_{f2}= -\kappa_{f3}= 2\kappa_{f4} =-2\kappa_{f1} \neq 0 \,, \nonumber\\
&&\
l_1 = l_2=l_3=l_4=l_5=0 \,, \qquad 
\mu_{f1}= n_1 = 0 \,.
\label{FRDS4}
\ea
Now we have the following four primary constraints:
\ba
	&&\calC^{(1)}_{\alpha_h}\equiv\pi_{\alpha_h} {\approx 0} \,, \qquad 
	\calC^{(1)}_{\alpha_f}\equiv\pi_{\alpha_f} {\approx 0} 	\,, \qquad 
	\calC^{(1)}_{\beta_h}\equiv\pi_{\beta_h} {\approx 0} \,, \qquad
	\calC^{(1)}_{\beta_f}\equiv\pi_{\beta_f} {\approx 0} \,. 
	\label{primary4p}
\ea
The total Hamiltonian can be expressed as
\ba
	\calH_T^S=\calH^S+\lambda_{\alpha_h}\calC^{(1)}_{\alpha_h}+\lambda_{\alpha_f}\calC^{(1)}_{\alpha_f}+\lambda_{\beta_h}\calC^{(1)}_{\beta_h}+\lambda_{\beta_f}\calC^{(1)}_{\beta_f}
	\,.
	\label{HT4p}
\ea
The evolution of the primary constraints is given by
\ba
	&& \calC^{(2)}_{\alpha_h}
		\equiv\{\calC^{(1)}_{\alpha_h},\calH_T^S\}
		=-8 (\mu_{h1} + \mu_{h2})\alpha_h
		+ 8(2k^2 \kappa_{h1} -3 \mu_{h2})\calR_h
		+8\mu_{h2} \calE_h
		{-4n_2(\alpha_f+3\calR_f-\calE_f )} \approx 0 \,, \label{4pC2ah}\\
	&& \calC^{(2)}_{\alpha_f}
		\equiv\{\calC^{(1)}_{\alpha_f},\calH_T^S\}
		=-8 \mu_{f2} \alpha_f 
		+ 8(2k^2 \kappa_{f1}-3\mu_{f2})\calR_f
		+ 8\mu_{f2} \calE_f
		{-4n_2(\alpha_h+3\calR_h -\calE_h)} \approx 0 \,, \label{4pC2af}\\
	&& \calC^{(2)}_{\beta_h}\equiv\{\calC^{(1)}_{\beta},\calH^S_T\}=-k\pi_{{\cal E}_h}+4\mu_{h1}\beta_h \approx 0 \,, \label{4pC2bh}\\
	&& {0 \approx} \calC^{(2)}_{\beta_f}\equiv\{\calC^{(1)}_{\beta_f},\calH^S_T\}=-k\pi_{{\cal E}_f} \approx 0 \,.\label{4pC2bf}
\ea
Here, all the secondary constraints cannot be trivially zero with any choice of the coefficients since $\kappa_{h1} \neq 0$ and $\kappa_{f1} \neq 0$.
First, let us take a look at the time evolution of the other primary constraints, that is, 
\ba
\begin{pmatrix}
		{\dot \calC}^{(2)}_{\alpha_{h}}\\
	{\dot \calC}^{(2)}_{\alpha_{f}}  \\
	{\dot \calC}^{(2)}_{\beta_{h}} 
\end{pmatrix}
=
\begin{pmatrix}
	\{ \calC^{(2)}_{\alpha_{h}} \,, {\calH^S_{T}} \}    \\
	\{ \calC^{(2)}_{\alpha_{f}} \,, {\calH^S_{T}} \}    \\
	\{ \calC^{(2)}_{\beta_{h}} \,, {\calH^S_{T}} \}  
\end{pmatrix}
=
\begin{pmatrix}
	\{ \calC^{(2)}_{\alpha_{h}} \,, {\calH^S} \}    \\
	\{ \calC^{(2)}_{\alpha_{f}} \,, {\calH^S} \}    \\
	\{ \calC^{(2)}_{\beta_{h}} \,, {\calH^S} \}  
\end{pmatrix}
+
\begin{pmatrix}
	-8(\mu_{h1} + \mu_{h2})&-4n_2&0\\
	-4n_2&-8\mu_{f2}&0\\
	0&0&4\mu_{h1}
\end{pmatrix}
\begin{pmatrix}
	\lambda_{\alpha_{h}}  \\
	\lambda_{\alpha_{f}}  \\
	\lambda_{\beta_{h}} 
\end{pmatrix}
\approx 0 \,.
\ea
When $n_2^2 -4\mu_{f2} (\mu_{h1}+\mu_{h2})\neq 0$, all the Lagrange multipliers, $\lambda_{\alpha_h}$, $\lambda_{\alpha_f}$,  and $\lambda_{\beta_h}$, are
determined by the above equations. 
As for $\calC^{(2)}_{\beta_f}$, it commutes with all the primary constraints and the consistency of $\calC^{(2)}_{\beta_f}$ gives the tertiary constraint,
\ba
	\calC^{(3)}_{\beta_f}=\{{\calC^{(2)}_{\beta_f}}
		,\calH_T^S\}
			=
			{-8k\mu_{f2}(\alpha_f+3\calR_f-\calE_f)-4kn_2(\alpha_h +3\calR_h -\calE_h )} \approx 0 \,.
\ea 
Now we redefine the secondary and tertiary constraints for $\beta_f$ as 
\ba
	\widetilde\calC^{(2)}_{\beta_f} 
&=&\calC^{(2)}_{\beta_f} -k\calC^{(1)}_{\alpha_f}
=-k (\pi_{\alpha_f} + \pi_{\calE_f}) \approx 0 \,. \\
	\widetilde\calC^{(3)}_{\beta_f}
		&=&\calC^{(3)}_{\beta_f}-k\calC^{(2)}_{\alpha_f}
		=-16k^3 \kappa_{f1} \calR_f \approx 0 \,.
\ea
Then, 
$\widetilde{\calC}^{(3)}_{\beta_f}$ does commute with $\calC^{(1)}_{\alpha_h}$ and $\calC^{(1)}_{\alpha_f}$, and one can see 
$\dot{\widetilde\calC}{}^{(3)}_{\beta_f}=k^2\calC^{(2)}_{\beta_f}\approx 0$, implying no more constraint is generated.
In addition, one can also check that the constraints, {$\calC^{(1)}_{\beta_f}$ and} $\widetilde\calC^{(2,3)}_{\beta_f}$, commute with all the constraints and
hence these are first class while the rest of constraints are second class. Therefore, we conclude the number of the physical DOFs is 
$(8\times 2 - 6 - 3  \times 2 )/2=2$ {when $n_2^2 -4\mu_{f2} (\mu_{h1}+\mu_{h2})\neq 0$}. 

In order to remove an extra DOF, we need to impose an additional constraint for the parameter :
\ba
	n_2^2 -4\mu_{f2} (\mu_{h1}+\mu_{h2})&=&0
	\label{const-n2}
	\,,
\ea
which yields two branches : 
\ba
	\mu_{h2} &=& -\mu_{h1} + \displaystyle{\frac{n_2^2}{4 \mu_{f2}}} \qquad {\rm (Class~II)} \,,
	\label{case2}\\
	\mu_{f2} &=& n_2 =0 \qquad \qquad \quad {\rm (Class~III)}
	\,.\label{case3}
	\label{DC:detM0}
\ea
Note that $\calC^{(3)}_{\beta_f}$ trivially vanishes in the second case (Class~III).

\subsubsection{Class II}
Let us consider Class II first. For convenience, we redefine the primary constraint for $\alpha_h$ with a linear combination of those for $\alpha_h$ and $\alpha_f$.
Then the four primary constrains reads :
\ba
&& \calC^{(1)}_{\alpha_h}\equiv\pi_{\alpha_h} - {n_2 \over 2\mu_{f2}} \pi_{\alpha_f} \approx 0 \,,\qquad 
\calC^{(1)}_{\alpha_f}\equiv\pi_{\alpha_f} \approx 0 \,, \qquad 
\calC^{(1)}_{\beta_h}\equiv\pi_{\beta_h} \approx 0 \,, \qquad
\calC^{(1)}_{\beta_f}\equiv\pi_{\beta_f} \approx 0 \,.
\ea
We have the same constrains from the evolution of the primary constraints for $\alpha_f$, $\beta_h$ and $\beta_f$ as in \eqref{4pC2af}, \eqref{4pC2bh} and \eqref{4pC2bf} respectively. 
Due to the condition \eqref{const-n2}, only one of the Lagrange multipliers, $\lambda_{\alpha_h}$ or $\lambda_{\alpha_f}$, is determined by the evolution of $\calC^{(2)}_{\alpha_h}$ or $\calC^{(2)}_{\alpha_f}$. 
Suppose that $\lambda_{\alpha_f}$ has been determined by the evolution of $\calC^{(2)}_{\alpha_f}$ though $\lambda_{\alpha_h}$ has not.
The evolution of the primary constraint for $\alpha_h$ demands
\ba
&&
\calC^{(2)}_{\alpha_h} 
\equiv\{\calC^{(1)}_{\alpha_h},\calH_T^S\}
=-8\mu_{h1} \calE_h -{8k^2 n_2 \kappa_{f1} \over \mu_{f2}}\calR_f
+ 8(2k^2\kappa_{h1} + 3\mu_{h1})\calR_h \approx 0 \,, \\
&& 
\calC^{(3)}_{\alpha_h}
\equiv\{\calC^{(2)}_{\alpha_h},\calH^S_T\}
= -8k\mu_{h1} \beta_h
-{k^2 n_2 \over 2 \mu_{f2}} \pi_{\calE_f}
+k^2 \pi_{\calE_h} - {\mu_{h1} \over 2\kappa_{h1}}\pi_{\calR_h} \approx 0 \,,
\ea
Since $\calC^{(3)}_{\alpha_h}$ does not commute with $\calC^{(1)}_{\beta_h}$, it is coveninent to introduce a linear combination
of $\calC^{(3)}_{\alpha_h}$ and $\calC^{(2)}_{\beta_h}$ as
\ba
	\widetilde\calC^{(3)}_{\alpha_h}=\calC^{(3)}_{\alpha_h}+2k \calC^{(2)}_{\beta_h}
	\,.
\ea
The evolution of $\widetilde\calC^{(3)}_{\alpha_h}$ yields the constraint $\widetilde\calC^{(4)}_{\alpha_h}=\{\widetilde\calC^{(3)}_{\alpha_h},\calH^S_T\} \approx 0$.
Since $\{\widetilde\calC^{(4)}_{\alpha_h},\calC^{(1)}_{\alpha_h}\} =-12 \mu_{h1}^2/\kappa_{h1}\neq 0$, 
the evolution of $\widetilde\calC^{(4)}_{\alpha_f}$ determines the Lagrange multiplier $\lambda_{\alpha_h}$ and no more constraint is generated.
It can be easily checked that $\widetilde\calC^{(3)}_{\alpha_h}$ and $\widetilde\calC^{(4)}_{\alpha_h}$ cannot be trivially zero. 
Since in this case
\ba
 \rm{first \, class} \quad &:& \quad \calC^{(1)}_{\beta_f} \,, \quad \widetilde\calC^{(2)}_{\beta_f} \,, \qquad \widetilde\calC^{(3)}_{\beta_f} \,,\\
 \rm{second \, class} \quad &:& \quad \calC^{(1)}_{\alpha_h} \,, \qquad \calC^{(2)}_{\alpha_h} \,, \qquad \widetilde\calC^{(3)}_{\alpha_h} \,, \qquad \widetilde\calC^{(4)}_{\alpha_f} \,, \qquad \calC^{(1)}_{\alpha_f} \,, \qquad \calC^{(2)}_{\alpha_f} \,, \qquad \calC^{(1)}_{\beta_h} \,, \qquad \calC^{(2)}_{\beta_h} \,,
\ea
 therefore we have : 
\ba
	\textrm{Scalar DOF} 
		&=&\frac{1}{2}\biggl[
			 8\times 2 - 8 \,(\textrm{3 primary \& 3 secondary \& 1 tertiary \& 1 quaternary\,)} 
	\notag\\
	&&\quad
			- 3 \,(\textrm{1 primary \& 1 secondary \& 1 tertiary)} \times 2 \,(\textrm{first-class})
			\biggr]
		=1 \,.
\ea
To summarize, we find another novel class of theory with a single DOF in the scalar sector :\\
{\bf Class II} :
	\ba
	&&\kappa_{h2}= -\kappa_{h3} = 2\kappa_{h4} =-2\kappa_{h1} \neq 0 \,, \qquad
	\kappa_{f2}= -\kappa_{f3}= 2\kappa_{f4} =-2\kappa_{f1} \neq 0 \,, \nonumber\\
	&&\
	l_1=l_2=l_3=l_4=l_5=0 \,, \qquad 
	\mu_{f1}=0 \,, \qquad
	n_1 = 0 \,, \qquad
	\mu_{h1}+ \mu_{h2} - {n_2^2 \over 4 \mu_{f2}} = 0 \,.
	\label{ClassII}
	\ea
The Lagrangian for class II is given by
\ba
{\cal L} &=&
-\left( \kappa_{h1} h_{\mu\nu}\widehat{\cal E}^{\mu\nu\alpha\beta}h_{\alpha\beta}+\kappa_{f1}f_{\mu\nu}\widehat{\cal E}^{\mu\nu\alpha\beta}f_{\alpha\beta}\right)
-\mu_{h1} (h_{\mu\nu}h^{\mu\nu}-h^2) -\frac{1}{4\mu_{f2}}\left( n_2h+2\mu_{f2}f\right)^2
\,.
\ea
One can check that this theory is invariant under the gauge transformation
\ba
h_{\mu\nu} &\to& {\widetilde h}_{\mu\nu} = 
h_{\mu\nu}\,,\\
f_{\mu\nu} &\to& {\widetilde f}_{\mu\nu} =
f_{\mu\nu} + \partial_\mu \xi_\nu+  \partial_\nu \xi_\mu \qquad {\rm with}\qquad\partial^\mu \xi_\mu=0 \,.
\ea
Again, due to the transverse condition in the gauge transformation, this theory is different from the linearized Hassan-Rosen bigravity.

\subsubsection{Class III}
In this case, we have the same primary constraints as well as the same Hamiltonian as before with the only exception that $\mu_{f2} = n_2=0$ and hence the subsequent constraints are the same.
To summarize, we have 
\ba
	\eqref{HT4p} \quad : \quad
	&& \calH_T^S=\calH^S+\lambda_{\alpha_h}\calC^{(1)}_{\alpha_h}+\lambda_{\alpha_f}\calC^{(1)}_{\alpha_f}+\lambda_{\beta_h}\calC^{(1)}_{\beta_h}+\lambda_{\beta_f}\calC^{(1)}_{\beta_f} \,,
\ea
 and 
\ba
	\eqref{primary4p} \quad : \quad
	&&
	\calC^{(1)}_{\alpha_h} \equiv\pi_{\alpha_h} \approx 0 \,, \quad 
	\calC^{(1)}_{\alpha_f}\equiv\pi_{\alpha_f} \approx 0 \,, \quad 
	\calC^{(1)}_{\beta_h}\equiv\pi_{\beta_h} \approx 0 \,, \quad
	\calC^{(1)}_{\beta_f}\equiv\pi_{\beta_f} \approx 0 \,. \\
	( \ref{4pC2ah} - \ref{4pC2bf}) \quad : \quad
	&& 
	\calC^{(2)}_{\alpha_h} \approx 0 \,, \qquad 
	\calC^{(2)}_{\alpha_f} \approx 0 \,, \qquad 
	\calC^{(2)}_{\beta_h} \approx 0 \,, \qquad 
	\calC^{(2)}_{\beta_f} \approx 0 \,.
\ea
The Lagrange multipliers $\lambda_{\alpha_h}$ and $\lambda_{\beta_h}$ are determined from the time-evolution of $\calC^{(2)}_{\alpha_h}$ and $\calC^{(2)}_{\beta_h}$. 
This is because $\{\calC^{(2)}_{\alpha_h},\calC^{(1)}_{\alpha_h}\} = -8(\mu_{h1} + \mu_{h2})$ and $\{\calC^{(2)}_{\beta_h},\calC^{(1)}_{\beta_h}\} = 4 \mu_{h1} \neq 0$
 with the fact that other Poisson brackets with the primary constraints vanish. 
In addition, the evolution of $\calC^{(2)}_{\alpha_f}$ and $\calC^{(2)}_{\beta_f}$ does not yield a new constraint since 
\ba
\dot{\calC}^{(2)}_{\alpha_f}= \{\calC^{(2)}_{\alpha_f},\calH^S_T\} = -k\calC^{(2)}_{\beta_f} \approx 0 \,,
 \ea
and $\dot{\calC}^{(2)}_{\beta_f}= \{\calC^{(2)}_{\beta_f},\calH^S_T\} \approx 0$. 
Therefore, we find two DOFs in the scalar sector:
\ba
\textrm{Scalar DOF} 
&=&\frac{1}{2}\biggl[
8\times 2 - 4 \,(\textrm{2 primary \& 2 secondary)} 
- 4 \,(\textrm{2 primary \& 2 secondary)} \times 2 \,(\textrm{first-class})
\biggr]
=2 \,, \notag\\
\ea
 since 
\ba
 \rm{first \, class} \quad &:& \quad \calC^{(1)}_{\alpha_f} \,, \qquad \calC^{(2)}_{\alpha_f} \,, \qquad \calC^{(1)}_{\beta_f} \,, \qquad \calC^{(2)}_{\beta_f} \,, \\
 \rm{second \, class} \quad &:& \quad \calC^{(1)}_{\alpha_h} \,, \qquad \calC^{(2)}_{\alpha_h} \,, \qquad \calC^{(1)}_{\beta_h} \,, \qquad \calC^{(2)}_{\beta_h} \,.
\ea

Now the only possible option to have a single DOF is to impose 
\ba
\mu_{h1} + \mu_{h2} = 0 \,,
\ea
 so that we obtain the tertiary constraint from $\alpha_h$.
In this case the tertiary constraint reads :
\ba
\calC^{(3)}_{\alpha_h}
=\{\calC^{(2)}_{\alpha_h},\calH^S_T\}
= -8k\mu_{h1} \beta_h +k^2 \pi_{\calE_h}  - {\mu_{h1} \pi_{\calR_h} \over 2\kappa_{h1}} \approx 0 \,.
\ea
Since $\calC^{(3)}_{\alpha_h}$ does not commute with $\calC^{(1)}_{\beta_h}$, let us define
\ba
\widetilde\calC^{(3)}_{\alpha_h}
=\calC^{(3)}_{\alpha_h} + 2k \calC^{(2)}_{\beta_h} \,.
\ea
The evolution of this constraint gives the quaternary constraint:
\ba
\widetilde\calC^{(4)}_{\alpha_h}
&=&\{\widetilde\calC^{(3)}_{\alpha_h},\calH^S_T\} =
{4 \mu_{h1} \over \kappa_{h1}} \left[
-3\mu_{h1} \alpha_h + 2 \mu_{h1} \calE_h + {2(k^2 \kappa_{h1}-3\mu_{h1})} \calR_h
\right] \approx 0 \,.
\ea
The time-evolution of $\widetilde\calC^{(4)}_{\alpha_h}$ determines the Lagrange multiplier $\lambda_{\alpha_h}$.
On the other hand the evolution of the secondary constraints for $\alpha_f$, $\beta_h$ and $\beta_f$ do not yield a new constraint.
The evolution of $\calC^{(2)}_{\alpha_f}$ and $\calC^{(2)}_{\beta_f}$ are trivial since $\dot{\calC}^{(2)}_{\alpha_f}= \{\calC^{(2)}_{\alpha_f},\calH^S_T\} = -k\calC^{(2)}_{\beta_f}$
and $\dot{\calC}^{(2)}_{\beta_f}= \{\calC^{(2)}_{\beta_f},\calH^S_T\} = 0$. 
The time-evolution of $\calC^{(2)}_{\beta_h}$ can be used to determine the Lagrange multiplier, $\lambda_{\beta_h}$.
Since
\ba
 \rm{first \, class} \quad &:& \quad \calC^{(1)}_{\alpha_f} \,, \qquad \calC^{(2)}_{\alpha_f} \,, \qquad \calC^{(1)}_{\beta_f} \,, \qquad \calC^{(2)}_{\beta_f} \,, \\
 \rm{second \, class} \quad &:& \quad \calC^{(1)}_{\alpha_h} \,, \qquad \calC^{(2)}_{\alpha_h} \,, \qquad \widetilde\calC^{(3)}_{\alpha_h} \,, \qquad \widetilde\calC^{(4)}_{\alpha_h} \,, \qquad \calC^{(1)}_{\beta_h} \,, \qquad \calC^{(2)}_{\beta_h} \,,
\ea
 we find
\ba
\textrm{Scalar DOF} 
&=&\frac{1}{2}\biggl[
8\times 2 - 6 \,(\textrm{2 primary \& 2 secondary \& 1 tertiary \& 1 quaternary\,)} 
\notag\\
&&\quad
- 4 \,(\textrm{2 primary \& 2 secondary)} \times 2 \,(\textrm{first-class})
\biggr]
=1 \,.
\ea
In this case
\\
{\bf Class III} :
	\ba
	&&\kappa_{h2}= -\kappa_{h3} = 2\kappa_{h4} =-2\kappa_{h1} \neq 0 \,, \qquad
	\kappa_{f2}= -\kappa_{f3}= 2\kappa_{f4} =-2\kappa_{f1} \neq 0 \,,\nonumber\\
	&&l_1=l_2=l_3 =  l_4 = l_5 =n_1=n_2=\mu_{f1}=\mu_{f2}=0 \,, \qquad 
	\mu_{h2}=-\mu_{h1} \,.
\label{ClassIII}
	\ea
The Lagrangian for class III is given by
	\ba
	{\cal L} &=&
	-\left( \kappa_{h1} h_{\mu\nu}\widehat{\cal E}^{\mu\nu\alpha\beta}h_{\alpha\beta}+\kappa_{f1}f_{\mu\nu}\widehat{\cal E}^{\mu\nu\alpha\beta}f_{\alpha\beta}\right)
	-\mu_{h1} (h_{\mu\nu}h^{\mu\nu}-h^2)
	\,.
	\ea
It is clear that this case corresponds to the linearized Hassan-Rosen bigravity, \eqref{HR1}.

\section{summary}
In this paper, we investigated a Lorentz invariant action for two rank-2 symmetric tensor fields $h_{\mu\nu}$ and $f_{\mu\nu}$. Based on the Hamiltonian analysis, we classified theories with seven physical degrees of freedom whose action consists of the most generic quadratic terms containing up to two derivatives with respect to spacetime for each term.
To simplify the problem, we have utilized a field redefinition to reduce the model parameter space. We then found three distinct classes of theories, which are not connected by a linear field redefinition. In any cases, the Hamiltonian structure in the tensor and vector sectors are the same, that is, one of the fields behaves as massless, and the other has non-vanishing mass in dispersion relations. The first theory, the Class I, contains three primary constraints in the scalar sector and is invariant under the transverse diffeomorphism. Furthermore, the kinetic terms for both fields do not take the form of Einstein-Hilbert term even by the field redefinition, and the mass term no longer has the Fierz-Pauli tuning. The Class II is also invariant under the transverse diffeomorphism but contains four primary constraints differently from the Class I. The kinetic terms for both fields are described by the Einstein-Hilbert terms, and a new tuning parameter enters in the mass matrix thanks to the transverse condition in the gauge transformation, which was absent in the linearized Hassan-Rosen bigravity. The Class III is nothing but the linearized Hassan-Rosen bigravity, which is invariant under the standard diffeomorphism. Since we have reduced the model parameter space by the linear field redefinition before the Hamiltonian analysis, a broader class of theories can be obtained by the field redefinition, which could be different theories depending on the matter coupling, although their Hamiltonian properties and physical degrees of freedom does not change. 

The transverse diffeomorphism appeared in the Class I and II can be nonlinearized by introducing the unimodular condition $\det g=1$, where $g$ is one of the metrics in bimetric gravity. Therefore, the first two class of theories, Class I and Class II, might open a new window of finding extended theories of massive bimetric gravity. In fact, if we linearize the Hassan-Rosen bigravity with the unimodular condition, one is able to obtain a part of the Class II, where all the mixing terms are switched off. Although such a case is trivial because the unimodular condition brings just a cosmological constant in the Einstein equation as the (massless) unimodular gravity, it would be interesting to investigate weather nonlinear completions of the Class II itself can be possible or not. Moreover, the nonlinearization of the Class I would be also interesting.

\acknowledgments
We would be grateful to Norihiro Tanahashi for the initial collaboration in the early stage of this work. 
A.N. would also like to thank Takahiro Tanaka for fruitful discussion and useful suggestion.
This work was supported in part by JSPS Grant-in-Aid for Scientific Research Nos.~JP17K14304 (D.Y.), JP19H01891 (A.N. and D.Y.) and 20H05852 (A.N.).

\appendix

\section{Linear field redefinition}
\label{sec:field redef}
In this appendix, we
 consider the transformation of the action for the fields $h_{\mu\nu}$ and $f_{\mu\nu}$ under a redefinition of them.
The most generic transformation linear in the fields is :
\ba
	&&h_{\mu\nu}=\Omega_h\overline h_{\mu\nu}+\omega_h \overline f_{\mu\nu}+\left(\Gamma_h\overline h+\gamma_h\overline f\right)\eta_{\mu\nu}
	\,,\\
	&&f_{\mu\nu}=\Omega_f\overline f_{\mu\nu}+\omega_f\overline h_{\mu\nu}+\left(\Gamma_f\overline f+\gamma_f\overline h\right)\eta_{\mu\nu}
	\,,
\ea
where $\Omega_{h,f}$ and $\Gamma_{h,f}$ are constants, $\overline h,\overline f$ are
the trace of $\overline h_{\mu\nu}\,,\overline f_{\mu\nu}$ contracted by $\eta_{\mu\nu}$.
Since $\Omega_h$ and $\Omega_f$ only change the normalization for each Lagrangian, 
we hereafter set $\Omega_h=\Omega_f=1$.
Applying the transformation to the generic action, one obtains :
\begin{align}
S &= \int \dd^4 x \Bigl(
- {\overline \calK}_h^{\alpha \beta | \mu \nu \rho \sigma} h_{\mu \nu, \alpha} h_{\rho \sigma, \beta}
- {\overline \calK}_f^{\alpha \beta | \mu \nu \rho \sigma} f_{\mu \nu, \alpha} f_{\rho \sigma, \beta}
-{\overline {\cal G}}^{\alpha \beta \mu \nu \rho \sigma} h_{\mu \nu, \alpha} f_{\rho \sigma, \beta}\notag\\
& \qquad \qquad \qquad
- {\overline \calM}_h^{\mu \nu \rho \sigma} h_{\mu \nu} h_{\rho \sigma} 
- {\overline \calM}_f^{\mu \nu \rho \sigma} f_{\mu \nu} f_{\rho \sigma} 
- {\overline \calN}^{\mu \nu \rho \sigma} h_{\mu \nu} f_{\rho \sigma} \Bigr) \,,
\end{align}
where the coefficients of the transformed Lagrangian read
\ba
	&&\overline\kappa_{h1}=\kappa_{h1}+\omega_f\left( l_1+\omega_f\kappa_{f1}\right)
	\,,\\
	&&\overline\kappa_{h2}=\kappa_{h2}+\omega_f\left( l_2+\omega_f\kappa_{f2}\right)
	\,,\\
	&&\overline\kappa_{h3}=2\Gamma_h\kappa_{h2}+\left( 1+4\Gamma_h\right)\kappa_{h3}
		+\omega_f\Bigl[ 2\gamma_f\kappa_{f2}+\left(\omega_f+4\gamma_f\right)\kappa_{f3}\Bigr]
	\notag\\
	&&\quad\quad\quad
		+\left(\gamma_f+\Gamma_h\omega_f\right) l_2+\left(\omega_f+4\gamma_f\right) l_3+\omega_f\left( 1+4\Gamma_h\right) l_5
	\,,\\
	&&\overline\kappa_{h4}=2\Gamma_h\left( 1+2\Gamma_h\right)\kappa_{h1}+\Gamma_h^2\kappa_{h2}
		+\Gamma_h\left(1+4\Gamma_h\right)\kappa_{h3}+\left(1+4\Gamma_h\right)^2\kappa_{h4}
	\notag\\
	&&\quad\quad\quad
		+2\gamma_f\left(\omega_f+2\gamma_f\right)\kappa_{f1}+\gamma_f^2\kappa_{f2}
		+\gamma_f\left(\omega_f+4\Gamma_f\right)\kappa_{f3}+\left(\omega_f+4\gamma_f\right)^2\kappa_{f4}
	\notag\\
	&&\quad\quad\quad
		+\Bigl[\gamma_f\left( 1+4\Gamma_h\right) +\omega_f\Gamma_h\Bigr]l_1+\gamma_f\Gamma_hl_2
		+\Gamma_h\left(\omega_f+4\gamma_f\right) l_3+\left(1+4\Gamma_h\right)\left(\omega_f+4\gamma_f\right) l_4
		+\gamma_f\left(1+4\Gamma_h\right) l_5
	\,,\\
	&&{\bar \mu}_{h1}=\mu_{h1}+\omega_f^2\mu_{f1}+\omega_fn_1
	\,,\\
	&&{\bar \mu}_{h2}=2\Gamma_h\left(1+2\Gamma_h\right)\mu_{h1}+\left(1+4\Gamma_h\right)^2\mu_{h2}
		+\Bigl[ 2\gamma_f\left(\omega_f+2\gamma_f\right)\mu_{f1}+\left(\omega_f+4\gamma_f\right)^2\mu_{f2}\Bigr]
	\notag\\
	&&\quad\quad\quad
		+\Bigl[\Gamma_h\omega_f+(1+4\Gamma_f)\gamma_f\Bigr] n_1
		+\left( 1+4\Gamma_h\right)\left(\omega_f+4\gamma_f\right) n_2
	\,,
\ea
and ${\bar \kappa}_{f1,f2,f3,f4}$ and ${\bar \mu}_{f1,f2}$ can be obtained by replacing the labels $h$ and $f$.
And also we find
\ba
	&&\overline l_1= \left(1+\omega_h\omega_f\right) l_1+2\omega_h\kappa_{h1}+2\omega_f\kappa_{f1}
	\,,\\
	&&\overline l_2= \left(1+\omega_h\omega_f\right) l_2+2\omega_h\kappa_{h2}+2\omega_f\kappa_{f2}
	\,,\\
	&&\overline l_3=\left(\Gamma_f +\omega_f\gamma_h\right)l_2+\left( 1+4\Gamma_f\right) l_3+\omega_f\left(\omega_h+4\gamma_h\right) l_5
	\notag\\
	&&\quad\quad\quad
		+2\gamma_h\kappa_{h2}+\left(\omega_h+4\gamma_h\right)\kappa_{h3}
		+2\omega_f\Gamma_f\kappa_{f2}+\omega_f\left( 1+4\Gamma_f\right)\kappa_{f3}
	\,,\\
	&&\overline l_4=\left(\Gamma_h+\Gamma_f+4\Gamma_h\Gamma_f+\omega_h\gamma_f+\omega_f\gamma_h+4\gamma_h\gamma_f\right) l_1
		+\left(\Gamma_h\Gamma_f +\gamma_h\gamma_f\right) l_2+\Bigl[\Gamma_h\left(1+4\Gamma_f\right)+\gamma_h\left(\omega_f+4\gamma_f\right)\Bigr] l_3
	\notag\\
	&&\quad\quad\quad
		+\Bigl[\left(1+4\Gamma_h\right)\left(1+4\Gamma_f\right)+\left(\omega_h+4\gamma_h\right)\left(\omega_f+4\gamma_f\right)\Bigr] l_4
		+\Bigl[\Gamma_f\left(1+4\Gamma_h\right)+\gamma_f\left(\omega_h+4\gamma_h\right)\Bigr] l_5
	\notag\\
	&&\quad\quad\quad
		+2\Bigl[\gamma_h\left( 1+4\Gamma_h\right) +\omega_h\Gamma_h\Bigr]\kappa_{h1}
		+2\gamma_h\Gamma_h\kappa_{h2}
		+\Bigl[\gamma_h\left( 1+8\Gamma_h\right) +\omega_h\Gamma_h\Bigr]\kappa_{h3}
		+2\left( 1+4\Gamma_h\right)\left(\omega_h+4\gamma_h\right)\kappa_{h4}
	\notag\\
	&&\quad\quad\quad
		+2\Bigl[\gamma_f\left( 1+4\Gamma_f\right) +\omega_f\Gamma_f\Bigr]\kappa_{f1}
		+2\gamma_f\Gamma_f\kappa_{f2}
		+\Bigl[\gamma_f\left( 1+8\Gamma_f\right) +\omega_f\Gamma_f\Bigr]\kappa_{f3}
		+2\left( 1+4\Gamma_f\right)\left(\omega_f+4\gamma_f\right)\kappa_{f4}
	\,,\\
	&&\overline l_5=\left(\Gamma_h+\omega_h\gamma_f\right) l_2+\omega_h\left(\omega_f+4\gamma_f\right) l_3+\left(1+4\Gamma_h\right) l_5
	\notag\\
	&&\quad\quad\quad
		+2\gamma_f\kappa_{f2}+\left(\omega_f+4\gamma_f\right)\kappa_{f3}
		+2\omega_h\Gamma_h\kappa_{h2}+\omega_h\left( 1+4\Gamma_h\right)\kappa_{h3}
	\,,\\
	&&\overline n_1=\left( 1+\omega_h\omega_f\right) n_1+2\omega_h\mu_{h1}+2\omega_f\mu_{f1}
	\,,\\
	&&\overline n_2=\left(\Gamma_h+\Gamma_f+4\Gamma_h\Gamma_f+\omega_h\gamma_f+\omega_f\gamma_h+4\gamma_h\gamma_f\right) n_1
		+\Bigl[\left(1+4\Gamma_h\right)\left(1+4\Gamma_f\right) +[\left(\omega_h+4\gamma_h\right)\left(\omega_f+4\gamma_f\right)\Bigr] n_2
	\notag\\
	&&\quad\quad\quad
		+2\Bigl[\gamma_h\left( 1+4\Gamma_h\right) +\omega_h\Gamma_h\Bigr]\mu_{h1}
		+2\left( 1+4\Gamma_h\right) \left(\omega_h+4\gamma_h\right) \mu_{h2}
	\notag\\
	&&\quad\quad\quad
		+2\Bigl[\gamma_f\left( 1+4\Gamma_f\right) +\omega_f\Gamma_f\Bigr]\mu_{f1}
		+2\left( 1+4\Gamma_f\right) \left(\omega_f+4\gamma_f\right) \mu_{f2}
	\,.
\ea

The inverse transformation of the fields is given by
\ba
{\overline f}_{\mu\nu} &=& \frac{1}{4} \left\{ 
 \frac{4 h_{\mu \nu} - 4 \omega_h f_{\mu \nu} - (h - \omega_h f) \eta_{\mu \nu}}{1 - \omega_h \omega_f} 
 + \frac{(1 + 4 \Gamma_f) h - (\omega_h+4 \gamma_h) f}{(1+ 4 \Gamma_h) (1 + 4 \Gamma_f) 
   - (\omega_f+4 \gamma_f) (\omega_h+4 \gamma_h)} \eta_{\mu \nu} \right\} \, \\
{\overline f}_{\mu\nu} &=& \frac{1}{4} \left\{ 
\frac{4 f_{\mu \nu} - 4 \omega_f h_{\mu \nu} - (f - \omega_f h) \eta_{\mu \nu}}{1 - \omega_h \omega_f}
+ \frac{(1 + 4 \Gamma_h) f - (\omega_f+4 \gamma_f) h}{(1 + 4 \Gamma_h) (1 + 4 \Gamma_f)
   - (\omega_f+4 \gamma_f) (\omega_h+4 \gamma_h)} \eta_{\mu \nu} \right\} \,,
\ea
 and that for the trace of the fields 
\ba
{\bar h} &=& \frac{(1+ 4 \Gamma_f) h - (\omega_h + 4\gamma_h)f}{1+4\Gamma_h + 4 \Gamma_f (1+4\Gamma_h) - (\omega_f + 4\gamma_f)(\omega_h + 4\gamma_h)} \,,\\
{\bar f} &=& \frac{(1+ 4 \Gamma_h) f - (\omega_f + 4\gamma_f)h}{1+4\Gamma_h + 4 \Gamma_f (1+4\Gamma_h) - (\omega_f + 4\gamma_f)(\omega_h + 4\gamma_h)} \,.
\ea
 where the inverse transformation exists only when 
\ba
1- \omega_h \omega_f &\neq& 0 \,, \label{NI1} \\
1+4\Gamma_h + 4 \Gamma_f (1+4\Gamma_h) - (\omega_f + 4\gamma_f)(\omega_h + 4\gamma_h) &\neq&0 \,.
\label{NI2}
\ea

\subsection{Transformation under vector conditions}
In this appendix, we show that one can impose \eqref{FRDV} by using the field redefinition, without loss of generality.
Here, we consider a specific field redefinition under the vector condition \eqref{ConditionV} to simplify the analysis.  Let us first consider the following transformation, 
\ba
	h_{\mu\nu}=\overline h_{\mu\nu}-\frac{l_1}{2\kappa_{h1}}\overline f_{\mu\nu}
	\,, \qquad
	f_{\mu\nu}=\overline f_{\mu\nu}
	\,,
\ea
and then one can find $\overline l_1=0$ in the transformed theories.
Moreover, in the case of $\kappa_{h3}\neq\kappa_{h1}$ and $\kappa_{f3}\neq\kappa_{f1}$, 
if one considers the field transformation defined as
\ba
	h_{\mu\nu}=\overline h_{\mu\nu}-\frac{2\kappa_{h1}-\kappa_{h3}}{2(\kappa_{h1}-\kappa_{h3})}\overline h \, \eta_{\mu\nu}
	\,, \qquad
	f_{\mu\nu}=\overline f_{\mu\nu}-\frac{2\kappa_{f1}-\kappa_{f3}}{2(\kappa_{f1}-\kappa_{f3})}\overline f \, \eta_{\mu\nu}
	\,,
\ea
one can transform to the theories with $\overline\kappa_{h3}=2\overline\kappa_{h1}$ and $\overline\kappa_{f3}=2\overline\kappa_{f1}$
with the use of the first two conditions of \eqref{ConditionV}. 
Next, when one considers the following transformation:
\ba
	&&h_{\mu\nu}=\overline h_{\mu\nu}+\left( -\frac{l_3}{\kappa_{h1}}\gamma_f \overline h+\gamma_h \overline f \right) \eta_{\mu\nu}
	\,,\label{eq:trsf1}\\
	&&f_{\mu\nu}=\overline f_{\mu\nu}+\left( -\frac{l_5}{\kappa_{f1}}\gamma_h \overline f+\gamma_f \overline h \right) \eta_{\mu\nu}
	\,,\label{eq:trsf2}
\ea
one can check that
the transformed parameters still satisfy the conditions: $\overline\kappa_{h3}=2\overline\kappa_{h1}$ and $\overline\kappa_{f3}=2\overline\kappa_{f1}$.
Under this conditions, using the transformaltion Eqs.~\eqref{eq:trsf1} and \eqref{eq:trsf2} in which only $\gamma_{f}$ is considered, one finds
\ba
	\overline l_3+\overline l_4+\overline l_5
	=l_3+l_4+l_5
			+\biggl[8\left(\kappa_{f1}+\kappa_{f4}\right) -\frac{l_3(l_3+4l_4+4l_5)}{\kappa_{h1}}\biggr]\gamma_f
	\,.
\ea
Hence, performing the transformation Eqs.~\eqref{eq:trsf1} and \eqref{eq:trsf2} with
\ba
	\gamma_f=-\frac{\kappa_{h1}(l_3+l_4+l_5)}{l_3(l_3+4l_4+4l_5)-8\kappa_{h1}(\kappa_{f1}+\kappa_{f4})}
	\,,\qquad 
	\gamma_h=0
	\,,
\ea
one can transform to the theories with $\overline\kappa_{h3}=2\overline\kappa_{h1}, \overline\kappa_{f3}=2\overline\kappa_{f1}$ and $\overline l_3+\overline l_4+\overline l_5=0$.

\subsection{$3$ primary case in the scalar sector}
\label{app:3primaryredef}
In this appedix, we show that the conditions \eqref{FRDS3} can be imposed by the field redefinition, without loss of generality. 
Let us first consider the first case of \eqref{S3Pb} for the original theory described by $h_{\mu\nu}$. 
In order to simply the Lagrangian, we impose 
\ba
&&\overline l_1=\overline l_4=0, \qquad \overline l_3 + \overline l_5=0, \qquad 
\overline\kappa_{h3}=2\overline\kappa_{h1}, \qquad \overline\kappa_{f3}=2\overline\kappa_{f1}, \qquad \overline n_1 =0 \,.
\label{S3Pimposed}
 \ea
These conditions determine the coefficients of the field redefinition as follows :
\ba
&&\omega_h = -{n_1 \over 2\mu_{h1}} \,, \qquad 
\omega_f = {n_1 \kappa_{h1} \over 2 \mu_{h1}\kappa_{f1}} \,,   \nonumber\\ 
&&\Gamma_h = -\frac{-2 l_3 \mu_{h1} (l_3 l_5 n_1 -2 (l_3+l_5)\mu_{h1} \kappa_{f1}) + l_3 n_1 (l_3n_1+2\mu_{h1}\kappa_{f1})\kappa_{h1} + n_1^2 \kappa_{f1} \kappa_{h1}^2}{8 \mu_{h1} (-l_3 n_1 + 2\mu_{h1}\kappa_{f1})(l_3 l_5 -\kappa_{f1} \kappa_{h1})} \,, \quad 
\Gamma_f = \frac{n_1 (2l_3 \mu_{h1} + n_1 \kappa_{h1})}{8\mu_{h1} (2\mu_{h1}\kappa_{f1}-l_3n_1)} \,,  \nonumber\\ 
&& \gamma_h = {n_1 \over 8\mu_{h1}} \,, \qquad
\gamma_f= - \frac{\kappa_{h1}}{\kappa_{f1}} \frac{l_3^2 l_5 n_1^2 + 2 l_3 \mu_{h1} \kappa_{f1}(2\mu_{h1}\kappa_{f1}-l_5n_1)+\kappa_{f1} (4l_5 \mu_{h1}^2 \kappa_{f1} + n_1 (l_5 n_1 + 2\mu_{h1} {\kappa_{f1}})\kappa_{h1})}{8 \mu_{h1} (l_3 n_1 {-} 2\mu_{h1}\kappa_{f1})(l_3 l_5 -\kappa_{f1} \kappa_{h1})} \,.
\label{TransS3P1}
\ea
Then the transformed Lagrangian satisfies
	\ba
&&\kappa_{h2}= -\kappa_{h3} = -2\kappa_{h1} \neq 0 \,, \qquad
\kappa_{f2}= -\kappa_{f3} =-2\kappa_{f1} \neq 0 \,, \qquad \kappa_{f4} = -\kappa_{f1} - \frac{3l_3^2}{8\kappa_{h1}}\nonumber\\
&&\
l_1 = l_2=l_4=\mu_{f1}=n_1=0 \,, \qquad 
l_5=-l_3 \,.
\label{S3P}
\ea
Here the bars are omitted. Thus the kinetic term for $f_{\mu\nu}$ is 
Einstein-Hilbert term, and all kinetic interactions between $h$ and $f$ are absent in this frame. 
Since these conditions are same as \eqref{FRDS3}, we conclude that the theories having the the first option of \eqref{S3Pb} are transformed to the theories with \eqref{FRDS3}.

Next let us consider the second case of \eqref{S3Pb}. Imposing the same conditions \eqref{S3Pimposed}, we find the transformation,
\ba
&&\omega_h = -{n_1 \over 2\mu_{h1}} \,, \qquad 
\omega_f = {n_1 \kappa_{h1} \over 2 \mu_{h1}\kappa_{f1}} \,, \nonumber\\ 
&& \Gamma_h = -\frac{4l_3l_5^2\mu_{h1}^2 + 2l_3\mu_{h1}(-l_5 n_1 + 2\mu_{h1}\kappa_{f1})\kappa_{h1} + n_1 ((l_3+l_5)n_1 + 2\mu_{h1} \kappa_{f1})\kappa_{h1}^2}{8 \mu_{h1} (2l_5 \mu_{h1} - n_1 \kappa_{h1})(l_3 l_5 -\kappa_{f1} \kappa_{h1})} \,, \qquad   
\Gamma_f = -{1 \over 4} \,, \nonumber\\
&&\gamma_h = \frac{l_5 n_1 + 2\mu_{h1} \kappa_{f1}}{8 l_5 \mu_{h1}-4n_1 \kappa_{h1}} \,, \quad
\gamma_f= - \frac{\kappa_{h1}}{\kappa_{f1}} \frac{2l_5^2\mu_{h1}(l_3 n_1 -2\mu_{h1}\kappa_{f1})-(l_5 (l_3+l_5)n_1^2 + 2l_5 \mu_{h1}n_1 \kappa_{f1}+4\mu_{h1}^2 \kappa_{f1}^2)\kappa_{h1}}{8 \mu_{h1} (2l_5 \mu_{h1} - n_1 \kappa_{h1})(l_3 l_5 -\kappa_{f1} \kappa_{h1})} \,, \notag\\
\label{TransS3P1}
\ea
 where the transformed Lagrangian satisfies \eqref{S3P} equivalent to \eqref{FRDS3}. 
 Therefore, both the first and second option of the $3$ primary case \eqref{S3Pb} can be mapped into  \eqref{FRDS3}.

\subsection{$4$ primary case in the scalar sector}
\label{app:4primaryredef}
In this appendix, we show that the conditions \eqref{FRDS4} can be imposed using the field redefinition without loss of generality.
Let us now consider the $4$ primary case \eqref{S4P} for the original theory described by $h_{\mu\nu}$. 
In order to simplify the Lagrangian, we here impose 
\ba
&&\overline l_1=0, \qquad 
\overline\kappa_{h3}=2\overline\kappa_{h1}, \qquad \overline\kappa_{f3}=2\overline\kappa_{f1},
\qquad 
\overline\kappa_{h4}=-\overline\kappa_{h1}, \qquad \overline\kappa_{f4}=-\overline\kappa_{f1},
\qquad
\overline \mu_{f1}=0 \,.
\label{S4Pimposed}
\ea
These conditions provides the field redefinition with the following coefficients : 
\ba
&&\omega_h = -{n_1 \over 2\mu_{h1}}, \qquad 
\omega_f = {n_1 \kappa_{h1} \over 2\mu_{h1} \kappa_{f1}} , \qquad 
\Gamma_h =-\frac{2l_3l_5 \mu_{h1} - l_3 n_1 \kappa_{h1}}{8 \mu_{h1} (l_3l_5 - \kappa_{f1} \kappa_{h1})} , \qquad 
\Gamma_f = -\frac{2l_3l_5 \mu_{h1} + l_5 n_1 \kappa_{h1}}{8 \mu_{h1} (l_3l_5 - \kappa_{f1} \kappa_{h1})}, \\ 
&&\gamma_h =\frac{l_3l_5 n_1 +2 l_3 \mu_{h1} \kappa_{f1}}{8 \mu_{h1} (l_3l_5 - \kappa_{f1} \kappa_{h1})}, \qquad
\gamma_f= -\frac{\kappa_{h1}(l_3l_5 n_1 - 2l_5 \kappa_{f1}\mu_{h1})}{8 \mu_{h1} \kappa_{f1}(l_3l_5 - \kappa_{f1} \kappa_{h1})} \,.
\ea
Then the transformed Lagrangian satisfies
\ba
&&\kappa_{h2}= -\kappa_{h3} = 2\kappa_{h4}= -2\kappa_{h1} \neq 0 \,, \qquad
\kappa_{f2}= -\kappa_{f3}= 2\kappa_{f4} =-2\kappa_{f1} \neq 0 \,, \nonumber\\
&&\
l_1 = l_2=l_3=l_4=l_5=0 \,, \qquad 
\mu_{f1}= n_1 = 0 \,.
\label{S4Pt}
\ea
 which is obviously equivalent to \eqref{FRDS4}.
Here we omit the bar of the coefficients. 
Thus in the $4$ primary case, we can always map them into two Einstein-Hilbert term with no kinetic interactions between $h$ and $f$.

\section{Lagrangian in the scalar sector}
\label{app:Lscalar}
The Lagrangian for $h$ scalar perturbations is given by
\ba
{\cal L}_{hh,\mathrm{kin}}^S &=&
4 (\kappa_{h1}+\kappa_{h2}+\kappa_{h3}+\kappa_{h4}) \dot{\alpha}_h^2
- (2 \kappa_{h1}+\kappa_{h2}) \dot\beta_h^2 
+ 12 (\kappa_{h1}+3 \kappa_{h4}) \dot{{\cal R}}_h^2
+ 4 (\kappa_{h1}+\kappa_{h4}) \dot{{\cal E}}_h^2 
\notag\\
&& 
- 4 (\kappa_{h3}+2 \kappa_{h4})\left( -3\dot{{\cal R}}_h + \dot{{\cal E}}_h 
\right) \dot{\alpha}_h
-8 (\kappa_{h1}+3 \kappa_{h4}) \dot{{\cal R}}_h \dot{{\cal E}}_h 
\,,\\
{\cal L}_{hh,\mathrm{cross}}^S &=&
-4
\left[  (\kappa_{h2}+\kappa_{h3}) \dot{\alpha}_h
+  (\kappa_{h2}+3 \kappa_{h3}) \dot{{\cal R}}_h
-  (\kappa_{h2}+\kappa_{h3}) \dot{\cal E}_h 
\right] k \beta_h \,,
\\
{\cal L}_{hh,\mathrm{mass}}^S &=&
-4 \Bigl[ k^2 (\kappa_{h1}+\kappa_{h4})+\mu_{h1}+\mu_{h2}\Bigr] \alpha_h^2
+ \Bigl[  k^2 (2 \kappa_{h1}+\kappa_{h2}) +2 \mu_{h1} \Bigr] \beta_h^2 
\notag\\
&& 
- 4\Bigl[ k^2 (3 \kappa_{h1}+\kappa_{h2}+3 \kappa_{h3}+9 \kappa_{h4})+3 (\mu_{h1}+3 \mu_{h2}) \Bigr] {\cal R}_h^2
- 4 \Bigl[ k^2 (\kappa_{h1}+\kappa_{h2}+\kappa_{h3}+\kappa_{h4})+\mu_{h1}+\mu_{h2} \Bigr] {\cal E}_h^2 
\notag\\
&& 
- 4\Bigl[ \Bigl( k^2 (\kappa_{h3}+6 \kappa_{h4})+6 \mu_{h2} \Bigr) {\cal R}_h
- \Bigl( k^2 (\kappa_{h3}+2 \kappa_{h4})+2 \mu_{h2} \Bigr) {\cal E}_h 
\Bigr] \alpha_h
\notag\\
&& 
+8 \Bigl[ k^2 (\kappa_{h1}+\kappa_{h2}+2 \kappa_{h3}+3 \kappa_{h4})+ (\mu_{h1}+3 \mu_{h2}) \Bigr] {\cal R}_h {\cal E}_h 
\,.
\ea
The Lagrangian for $f$ perturbations can be obtained by replacing the above Lagrangian for $h$ with $f$. 
\ba
\calL^S_{hf,{\rm kin}}&=&4\left( l_2+l_3+l_4+l_5\right)\dot\alpha_h\dot\alpha_f-l_2\dot\beta_h\dot\beta_f+36l_4\dot{\cal R}_h\dot{\cal R}_f+4l_4\dot{\cal E}_h\dot{\cal E}_f
+12\left( l_3+l_4\right)\dot\alpha_h\dot{\cal R}_f+12\left( l_4+l_5\right)\dot\alpha_f\dot{\cal R}_h
\notag\\
&&-4\left( l_3+l_4\right)\dot\alpha_h\dot{\cal E}_f-4\left( l_4+l_5\right)\dot\alpha_f\dot{\cal E}_h-12l_4\left(\dot{\cal R}_h\dot{\cal E}_f+\dot{\cal R}_f\dot{\cal E}_h\right)
\,,\\
\calL^S_{hf,{\rm cross}}&=&-2\left[\left( l_2+2l_5\right)\dot\alpha_h+\left( l_2+6l_5\right)\dot{\cal R}_h-\left( l_2+2l_5\right)\dot\calE_h\right] k\beta_f
\notag\\
&&
-2\left[\left( l_2+2l_3\right)\dot\alpha_f+\left( l_2+6l_3\right)\dot{\cal R}_f-\left( l_2+2l_3\right)\dot\calE_f\right] k\beta_h
\,,\\
\calL^S_{hf,{\rm mass}}&=&-4\left( k^2l_4+n_1+n_2\right)\alpha_h\alpha_f+\left( k^2l_2+2n_1\right)\beta_h\beta_f
-4\left[ k^2(l_2+3l_3+9l_4+3l_5)+3n_1+9n_2\right]{\cal R}_h{\cal R}_f
\notag\\
&&-4\left[ k^2(l_2+l_3+l_4+l_5) +n_1+n_2\right]{\cal E}_h{\cal E}_f
-4\left[ k^2(3l_4+l_5)+3n_2\right)\alpha_h{\cal R}_f-4\left( k^2(l_3+3l_4)+3n_2\right]\alpha_f{\cal R}_h
\notag\\
&&+4\left[ k^2(l_4+l_5)+n_2\right]\alpha_h{\cal E}_f+4\left[ k^2(l_3+l_4)+n_2\right]\alpha_f{\cal E}_h
+4\left[ k^2(l_2+l_3+3l_4+3l_5)+n_1+3n_2\right]{\cal R}_h{\cal E}_f
\notag\\
&&+4\left[ k^2(l_2+3l_3+3l_4+l_5)+n_1+3n_2\right]{\cal R}_f{\cal E}_h
\,.
\ea

\section{2 primary case in vector sector}
\label{app:v2p}
In this Appendix, we investigate the Hamiltonian analysis in the case of $2$ primary constraints in the vector sector, where 
\ba
\kappa_{f2} = -2\kappa_{f1} + \frac{l_2^2}{4 (2\kappa_{h1}+\kappa_{h2})} \,,
\ea
is satisfied. In this case, 
we have the following two primary constraints, which is defined by
\ba
\calC^{(1)}_{\Bfi} \equiv \piBfi - \frac{l_2}{2(2\kappa_{h1} + \kappa_{h2})} \piBhi\approx 0 \,.
\ea
Then we define the total Hamiltonian
\ba
\calH^V_{T} = \calH^V + \lambda_{\Bfi} \calC^{(1)}_{\Bfi} \,.
\ea
The consistency of the primary constraints gives the secondary constraints 
\ba
\calC^{(2)}_{\Bfi}  &\equiv& \{ \calC^{(2)}_{\Bfi} \,, \calH^V_{T} \}  
={\left( 4\mu_{f1}-\frac{{l_2}n_1}{2\kappa_{h1}+\kappa_{h2}}\right)\Bfi
+2\left( n_1-\frac{{l_2}\mu_{h1}}{2\kappa_{h1}+\kappa_{h2}}\right)\Bhi
-k\piFfi +\frac{kl_2}{2(2\kappa_{h1}+\kappa_{h2})}\piFhi}
\,,
\ea
and the time-evolution of the secondary constraints gives
\ba
\calC^{(3)}_{\Bfi}  &\equiv& \{ \calC^{(2)}_{\Bfi} \,, \calH^V_{T} \}  
=\{ \calC^{(2)}_{\Bfi} \,, \calH^V \}  + \lambda_{\Bfi} \{ \calC^{(2)}_{\Bfi} \,,\calC^{(1)}_{\Bfi} \}  
\approx0\,,
\ea
where
\ba
\{ \calC^{(2)}_{\Bfi} \,,\calC^{(1)}_{\Bfi} \}  
= 4\mu_{f1} + \frac{l_2(l_2\mu_{h1}-2n_1(2\kappa_{h1}+\kappa_{h2}))}{(2\kappa_{h1}+\kappa_{h2})^2}\,.
\ea
Therefore, when $\{ \calC^{(2)}_{\Bfi} \,,\calC^{(1)}_{\Bfi} \}   \neq 0$, 
the Lagrange multipliers $ \lambda_{\Bfi}$ are determined by the above equation, and the primary and secondary constraints are second class. 
Therefore, the number of the physical DOFs in the vector sector is $(8\times 2 -4)/2 =6$. 

In order to further reduce the variable in the phase space, we need to impose an extra condition. 
The only option here is $\{ \calC^{(2)}_{\Bfi} \,,\calC^{(1)}_{\Bfi} \}  = 0$,
i.e.,
\ba
\mu_{f1}= -\frac{l_2(l_2\mu_{h1}-2n_1(2\kappa_{h1}+ \kappa_{h2}))}{4(2\kappa_{h1}+\kappa_{h2})^2}\,.
\ea
Then,
$\calC^{(3)}_{\Bfi}$ serves as the tertiary constraints, 
\ba
\calC^{(3)}_{\Bfi} &=& \frac{\left[
	n_1(2\kappa_{h1}+ \kappa_{h2}) -l_2 \mu_{h1}	
	\right]
	\left[
	{k}l_2 \Ffi + 2k (2\kappa_{h1}+\kappa_{h2})\Fhi
	-\piBhi
	\right]}{(2\kappa_{h1}+\kappa_{h2})^2} \approx 0 \,,
\ea
and, since $\{\calC^{(3)}_{\Bfi},\calC^{(1)}_{\Bfi}\}=0$, subsequently we have quaternary constraints,
\ba
\calC^{(4)}_{\Bfi} &\equiv& \{ \calC^{(3)}_{\Bfi} \,, \calH^V_{T} \}  = - \frac{\left[
	n_1(2\kappa_{h1}+ \kappa_{h2}) -l_2 \mu_{h1}	
	\right]
	\left[
	\left({k^2}l_2 +2n_1\right) \Bfi + 2\left( k^2 (2\kappa_{h1}+\kappa_{h2})+2\mu_{h1}\right)\Bhi
	-k\piFhi
	\right]}{(2\kappa_{h1}+\kappa_{h2})^2} \approx 0 \,. \nonumber\\
\ea
Now the time-evolution of the quaternary constraints gives
\ba
{\dot\calC^{(4)}_{\Bfi}}=\{ \calC^{(4)}_{\Bfi} \,, \calH^V_{T} \}  
=\{ \calC^{(4)}_{\Bfi} \,, \calH^V \}  + \lambda_{B_{f,i}} \{ \calC^{(4)}_{B_{f,i}} \,,\calC^{(1)}_{\Bfi} \} \approx0\,,
\ea
where
\ba
\{\calC^{(4)}_{\Bfi} \,,\calC^{(1)}_{\Bfi} \} 
= -\frac{	2\left(n_1(2\kappa_{h1}+ \kappa_{h2}) -l_2 \mu_{h1}\right)^2	}{(2\kappa_{h1}+\kappa_{h2})^3}\,.
\ea
Therefore, as long as $\{\calC^{(4)}_{\Bfi} \,,\calC^{(1)}_{\Bfi} \} $ is non-vanishing, the Lagrange multiplier are determined by the above equations, and the number of the physical DOFs is $(8 \times 2 -8)/2=4$. 

When $n_1(2\kappa_{h1}+ \kappa_{h2}) -l_2 \mu_{h1}=0$, 
the above tertiary constraint trivially vanishes and its time-evolution does not generate the independent constraint. 
In this case, we only have the primary and secondary constraints, but now they are first class since all the primary and secondary commute each other. 
Thus, the number of the physical DOFs is $(8\times 2 -4 \times 2)/2 = 4$. 
Therefore, the case with $2$ primary constraints in the vector sector cannot have $2$ physical DOFs.

\section{2 primary case : ${\rm det} \, {\cal K}^S\neq 0$ in scalar sector}
\label{app:twoprimary}

Let us consider the case with two primary constraints, namely the degenerate condition for the scalar components is not imposed, and the parameters only satisfies the vector conditions \eqref{ConditionV}.
We define the following two primary constraints for convenience,
\ba
\calC^{(1)}_{\beta_h}\equiv \pi_{\beta_h} \approx 0 \,, \qquad 
\calC^{(2)}_{\beta_f}\equiv\pi_{\beta_f}-\frac{n_1}{2\mu_{h1}}\pi_{\beta_h} \approx 0 \,.
\ea
The total Hamiltonian is defined as 
\ba
\calH_T^S=\calH+\lambda_{\beta_h}\calC^{(1)}_{\beta_h}+\lambda_{\beta_f}\calC^{(1)}_{\beta_f}
\,.
\ea
The evolution of the two primary constraints yields two secondary constraints:
\ba
&&\calC^{(2)}_{\beta_h}
=\{\calC^{(1)}_{\beta_h},\calH_T^S\}
=-k\left(\pi_{\alpha_h}+\pi_{{\cal E}_h}\right)+4\mu_{h1}\beta_h +2n_1\beta_f \approx 0 \,,\\
&&\calC^{(2)}_{\beta_f}
=\{\widetilde\calC^{(1)}_{\beta_f},\calH_T^S\}
=-k(\pi_{\alpha_f}+\pi_{{\cal E}_f})
+\frac{k n_1}{2\mu_{h1}}
(\pi_{\alpha_h}+\pi_{{\cal E}_h}) \approx 0 \,.
\ea
Since $\{\calC^{(2)}_{\beta_h},\calC^{(1)}_{\beta_h}\} =4\mu_{h1}\neq 0$, the Lagrange multiplier $\lambda_{\beta_h}$ is 
determined by imposing $\dot\calC^{(2)}_{\beta_h}\approx 0$, namely $\lambda_{\beta_h}\approx-\{\calC^{(2)}_{\beta_h},\calH\}/\{\calC^{(2)}_{\beta_h},\calC^{(1)}_{\beta_h}\}$.
The evolution of the remaining secondary constraint yields the tertiary constraint:
\ba
&&\calC^{(3)}_{\beta_f}
\equiv\{\calC^{(2)}_{\beta_f},\calH_T^S\}
={2k^3\Biggl[
-2l_5\alpha_h + \frac{ l_3 n_1}{\mu_{h1}} \alpha_f
- 2 \left(3 l_5 - \frac{2n_1 \kappa_{h1}}{\mu_{h1}}\right)\calR_h
-\left(8 \kappa_{f1} - \frac{3l_3 n_1}{\mu_{h1}}\right)\calR_f
+2l_5\calE_h - \frac{ l_3 n_1}{\mu_{h1}}\calE_f
\Biggr]} \approx 0 \,.\notag\\
\ea
Here, the tertiary constraint cannot be trivially zero since $\kappa_{f1} \neq 0$.
One can also check that $\dot\calC^{(3)}_{\beta_f}=k^2\calC^{(2)}_{\beta_f}\approx 0$, implying no more constraint is generated. 
The constraints $\calC^{(1,2,3)}_{\beta_f}$ commute with all other constraints, and therefore, we have three first-class constraints $\calC^{(1,2,3)}_{\beta_f}$ and two second-class constraints $\calC^{(1,2)}_{\beta_h}$.
Hence, the number of the physical DOFs is $(8\times 2 -2 -3\times 2)/2=4$. Since there is no further option to eliminate DOFs, one cannot obtain $1$ DOF theory in this case.

\section{Explicit expression of constraints}
\label{app:constraints}

In this appendix, we give explicit expression of the constraints in the case of the Class I. 
Eq.~\eqref{Caf3} is given by
\ba
\calC^{(3)}_{\alpha_f}&\equiv&\{\calC^{(2)}_{\alpha_f},\calH_T^S\} =
-\frac{8kl_3 \mu_{h1}}{\kappa_{h1}}\beta_h 
+c_3^{\alpha_h}\pi_{\alpha_h}
+ c_3^{\calR_h}\pi_{\calR_h}
+ c_3^{\calR_f}\pi_{\calR_f}
+ {k^2 l_3 \over \kappa_{h1}} \pi_{\calE_h}
+ k^2 \pi_{\calE_f} \approx 0 \,,
\ea
where 
\ba
c_3^{\alpha_h} &=&
\frac{1}{4\kappa_{h1}}\biggl[
l_3\left( k^2+\frac{6\mu_{h1}}{\kappa_{h1}}\right){-}(16\kappa_{h1}\kappa_{f1}+9l_3^2)y
\biggr]
\,,\\
c_3^{\calR_h} &=& 
\frac{l_3}{4\kappa_{h1}}\left(
k^2-\frac{2\mu_{h1}}{\kappa_{h1}}+3l_3y
\right)
\,,\\
c_3^{\calR_f} &=&-l_3y\,,
\ea
and
\ba
	y=\frac{2n_2\kappa_{h1}+3l_3\mu_{h1}}{{2}\kappa_{h1}(8\kappa_{f1}(\kappa_{h1}+\kappa_{h4})+3l_3^2)}
	\,.
\ea
After rescaling, Eq.~\eqref{Caf34b} is given by
\ba
\widetilde\calC^{(3)}_{\alpha_f}&\equiv&\{\calC^{(2)}_{\alpha_f},\calH_T^S\} =
\widetilde c_3^{\alpha_h}\pi_{\alpha_h}
+\widetilde c_3^{\calR_h}\pi_{\calR_h}
+c_3^{\calR_f}\pi_{\calR_f}
-{k^2 l_3 \over \kappa_{h1}} \pi_{\calE_h} \approx 0 \,,
\ea
where
\ba
	&&\widetilde c_3^{\alpha_h}=c_3^{\alpha_h}-\frac{5k^2l_3}{4\kappa_{h1}}
		=\frac{1}{4\kappa_{h1}}\biggl[
l_3\left( -{4}k^2+\frac{6\mu_{h1}}{\kappa_{h1}}\right){-}(16\kappa_{h1}\kappa_{f1}+9l_3^2)y
\biggr]
	\,,\\
	&&\widetilde c_3^{\calR_h}=c_3^{\calR_h}-\frac{k^2l_3}{4\kappa_{h1}}
		=\frac{l_3}{4\kappa_{h1}}\left(-\frac{2\mu_{h1}}{\kappa_{h1}}+3l_3y\right)
	\,.
\ea
The time evolution of $\widetilde\calC^{(3)}_{\alpha_f}$ yields the constraint
\ba
	\dot{\widetilde\calC}{}^{(3)}_{\alpha_f}
		\equiv \widetilde\calC^{(4)}_{\alpha_f}
		=\widetilde c_4^{\alpha_h}\alpha_h+\widetilde c_4^{\alpha_f}\alpha_f
			+\widetilde c_4^{\calE_h}\calE_h+\widetilde c_4^{\calE_f}\calE_f
			+\widetilde c_4^{\calR_h}\calR_h+\widetilde c_4^{\calR_f}\calR_f \approx 0 \,,
\ea
where
\ba
	&&\widetilde c_4^{\alpha_h}
		=8\left( 4\kappa_{f1}(\kappa_{h1}+\kappa_{h4})+l_3^2\right) k^2y
			+2\left(\frac{3l_3}{\kappa_{h1}}(2\kappa_{h1}n_2+3l_3\mu_{h1})+16\kappa_{f1}(\mu_{h1}+\mu_{h2})\right) y
			-\frac{12l_3\mu_{h1}^2}{\kappa_{h1}^2}
	\,,\\
	&&\widetilde c_4^{\alpha_f}
		=\frac{1}{\kappa_{h1}^2}\biggl[
				-2l_3(4\kappa_{h1}\kappa_{f1}+3l_3^2)(2\kappa_{h1}k^2+3\mu_{h1})y
				+8\kappa_{f1}(2\kappa_{h1}n_2+3l_3\mu_{h1})y-2k^2l_3^2(2k^2\kappa_{h1}-\mu_{h1})
			\biggr]
	\,,\\
	&&\widetilde c_4^{\calE_h}
		=-8\left( 4\kappa_{f1}(\kappa_{h1}+\kappa_{h4})+l_3^2\right) k^2y
			+2\left(\frac{3l_3}{\kappa_{h1}}(-2\kappa_{n1}n_2+l_3\mu_{h1})-16\kappa_{f1}\mu_{h2}\right)y+\frac{4l_3(2k^2\kappa_{h1}-\mu_{h1})\mu_{h1}}{\kappa_{h1}^2}
	\,,\\
	&&\widetilde c_4^{\calE_f}
		=\frac{4l_3(4\kappa_{h1}\kappa_{f1}+3l_3^2)}{\kappa_{h1}}k^2y+2\left( -8\kappa_{f1}n_2+\frac{9l_3^3\mu_{h1}}{\kappa_{h1}^2}\right)y
			+\frac{2k^2l_3^2(2k^2\kappa_{h1}-\mu_{h1})}{\kappa_{h1}^2}
	\,,\\
	&&\widetilde c_4^{\calR_h}
		=8\left( 4\kappa_{f1}(\kappa_{h1}+3\kappa_{}h4)-4l_3^2\right) k^2y+2\left(\frac{9l_3}{\kappa_{h1}}(2\kappa_{h1}n_2-l_3\mu_{h1})+48\kappa_{f1}\mu_{h2}\right) y \notag\\
		&&~~~~~~~~ -\frac{4l_3(4k^4\kappa_{h1}^2-2k^2\kappa_{h1}\mu_{h1}-3\mu_{h1}^2)}{\kappa_{h1}^2}
	\,,\\
	&&\widetilde c_4^{\calR_f}
		=-\frac{4l_3(8\kappa_{h1}\kappa_{f1}+9l_3^2)}{\kappa_{h1}}k^2y+6\left( 8\kappa_{f2}n_2-\frac{9l_3^2\mu_{h1}}{\kappa_{h1}^2}\right) y
			-\frac{6k^2l_3^2(2k^2\kappa_{h1}-\mu_{h1})}{\kappa_{h1}^2}
	\,.
\ea
Using other constraints, we then rewrite $\widetilde\calC^{(4)}_{\alpha_f}$ as
\ba
	\widetilde\calC^{(4)}_{\alpha_f}
		=&&\frac{2n_2\kappa_{h1}+3l_3\mu_{h1}}{\kappa_{h1}(8\kappa_{f1}(\kappa_{h1}+\kappa_{h4})+3l_3^2)}
			\Biggl[
				\frac{l_3}{2}\calC^{(2)}_{\alpha_f}
				-\frac{6l_3^3\mu_{h1}^2+\kappa_{f1}^2\kappa_{f1}(k^2(2\kappa_{h1}n_2+3l_3\mu_{h1})-4n_2\mu_{h1})}{3kl_3^2\mu_{h1}^2}\calC^{(3)}_{\beta_f}
	\notag\\
			&&\quad -16\kappa_{f1}\left(\mu_{h2}+\frac{\kappa_{h1}^2n_2^2}{3l_3^2\mu_{h1}}+\frac{1}{4}\mu_{h1}\right)
					\left({\cal E}_h-\alpha_h-3{\cal R}_h\right)
			\Biggr]
	\notag\\
			&&\quad +\frac{1}{\kappa_{h1}^2(2k^2\kappa_{h1}+3\mu_{h1})}
				\left( 1-\frac{\kappa_{h1}\kappa_{f1}(2n_2\kappa_{h1}+3l_3\mu_{h1})^2}{3l_3^2\mu_{h1}^2(8\kappa_{f1}(\kappa_{h1}+\kappa_{h4})+3l_3^2)}\right)
				\biggl[
						k\kappa_{h1}^2\left(2k^2\kappa_{h1}-\mu_{h1}\right)\calC^{(3)}_{\beta_f}
	\notag\\
			&&\quad\quad
						+4\mu_{h1}\left(4k^2\kappa_{h1}^2n_2+3l_3\mu_{h1}^2\right)\left({\cal E}_h-\alpha_h-3{\cal R}_h\right)
						-48l_3\mu_{h1}^2\left(2k^2\kappa_{h1}{\cal R}_h-\mu_{h1}\alpha_h\right)
					\biggr]
	\,.\label{eq:C^4_alpha_f}
\ea
When we further impose the two additional conditions \eqref{LCon:3primary} and \eqref{classIb}, the right-hand-side of the above equation reduces to 
the linear combination of the constraints, namely the time evolution of $\widetilde\calC^{(3)}_{\alpha_f}$ becomes trivial.

\bibliography{references}

\end{document}